\documentclass[11pt]{article}
\usepackage{amssymb}
\usepackage{amscd}
\usepackage[dvips]{graphicx}
\usepackage{yfonts}
\usepackage{txfonts}
 \linethickness{0.4pt}

\begin{document}
\newtheorem{lemma}{Lemma}[section]
\newtheorem{theorem}{Theorem}[section]
\newtheorem{corollary}{Corollary}[section]
\newtheorem{definition}{Definition}[section]
\newtheorem{proposition}{Proposition}[section]
\newcommand{\tot}{{\rm {tot}}}
\newcommand{\rel}{{\rm{rel}}}
        

\bigskip
\bigskip
\begin{center}
\Large{\bf{ Cohomological resolutions for anomalous Lie constraints }}
\end{center}
\bigskip
\bigskip
\begin{center}
{\large\bf Zbigniew Hasiewicz, Cezary J. Walczyk  }
\end{center}
\bigskip
\begin{center}
{
{Department of Physics, University of Bia{\l}ystok,\\
ul. Lipowa 41, 15-424 Bia{\l}ystok, Poland}}
\end{center}
\bigskip\bigskip

\date{June 2013}

\begin{abstract}
\noindent {It is shown that the BRST resolution of the spaces of
physical states of the systems with anomalies
can be consistently defined. The appropriate anomalous complexes are
obtained by canonical restrictions of the ghost extended spaces to
the kernel of anomaly operator without any modifications of the "matter" sector. The cohomologies of the  anomalous complex for the case of constarints constituting a centrally extended simple Lie algebra of compact type are calculated and analyzed in details within the framework of Hodge - deRham - K\"{a}hler theory: the vanishing theorem of the relative cohomologies is proved and the absolute cohomologies are reconstructed.}
\end{abstract}
\vspace{1.5cm}
\thispagestyle{empty}
\eject
\section*{Introduction}
The cohomological approach to contrained systems or systems with gauge symmetries was initiated
in the seventies of the last century \cite{b_r_s_t}. Since that time it had grown into quite advanced and  powerful machinery with successful applications in  field theory as well as in string theory \cite{strings_fields}. It is still under investigation and development on the classical and quantum levels. The cohomological BRST formalizm  appeared to be very efficient tool to describe the interactions of fields and/or strings.\\ 

The BRST cohomological approach is well established for first class \cite{Dirac} system of constraints. Most of the interesting physical systems are governed by the constraints of mixed type. \\
Its generalization to the systems of mixed type is not unique and there are several approaches. One of the proposals is to solve all costraints of second class already on the classical level in order to obtain the first class classical system to be quantized. This approach has two important drawbacks. 
First of all it might appear that upon quantization the first class system gets quantum anomaly (which happens mainly in the case of infinitely many degrees of freedom) as in field theory or string theory. Secondly one might obtain completely inadequate picture of the system at quantum level. The simplest example which comes into mind is a particle interacting with a centrally symmetric potential (eg. hydrogen atom) with constraint which fixes one of its angular momentum at non zero value. The reduction of this system on the classical level leads after canonical quantization to "flat" picture of its states and to wrong spectrum of agular momentum upon quantization.  The appropriate reduction on the quantum level based on the Gupta - Bleuler \cite{gb} polarization of connstraints leads to different and consistent result. The reamark  based on this simple example leads one to the conclusion that the diagram:
\[
\begin{CD}
\cal{C} @>{\mathrm{quantization}}>>  @. \cal{Q}\\
@V{\mathrm{classical\; reduction}}VV   @. @VV{\mathrm{GB\; quantum\; reduction}}V\\
\cal{C}_{\mathrm{red}} @>{\mathrm{quantization}}>> \cal{Q}_{\mathrm{red}}'\stackrel{?}{\leftrightarrow} @.\cal{Q}_{\mathrm{red}}
\end{CD}
\]
can not be coverted into commutative one, as it seems that an appropriate map marked by $?$ cannot be cosistently defined
\footnote{The vertical arrow on the left hand side of the diagram is strictly defined \cite{m_w} witin the framework of symplectic geometry}.
 What is most imortant in the above example: the way of proceeding according to Gupta-Bleuler rules \cite{gb}, indicated by the right hand side of the  diagram gives the  physically acceptable result in agreement with common intuition and knowledge. One may also think about constrained quantum system without any relation of unerlying cassical one, as it happened with Dual Theory \cite{jacob} based on the axiomatic approach to S-matrix.  The next examples, which evidently indicate that the diagram above cannot be converted into commutative one are given by the models of critical massive strings and non critical massless strings \cite{hasjas}. \\ 
For this reason the approach based on the polarization of the quantum constraints, which allows one to proceed  with an equivalent system of first class  at the quqntum level seems to be reasonable. \\
The situation is more or less standard if the algebra of constraints admits a real polarization - which is rather rarely encountered case within the class of the physical systems of importance. The problem becomes far from obvious when the polarization is neccessarily complex.  This last case includes the  most important physical theories and models: quantum electrodynamics \cite{gb}, non-critical string theories \cite{hasjas} and high spin systems \cite{my}.\\
There were some early proposals how to treat the constrained system in this situation \cite{GB} but that approach was prematured, far from being canonical and cosistent. \\
The canonical and mathematically consistent approach to cohomological BRST description of constrained systems of mixed class is  proposed in this paper. Although it was grown on the backgrounds of the experience, in string theory \cite{my_string} and high spin systems \cite{my}, the authors are convinced that  it's main ideas and results are universal as the underlying constructions can be easily (neglecting technical difficulties) extended to the wide class of models of physical importance. For this reason and in order to avoid technical difficulties, wchich would screen the main ideas, the authors decided to restrict the considerations to the case of constraints based on simple Lie algebra (avoiding algebraic coplications) of compact type (avoiding analytical complications) with trivial (which is implied by previous assumptions) and regular \cite{gs} anomaly (again, in order to avoid algebraic complexity). \\
\\
The paper is organized as follows. In the first chapter the differential space which contains the anomalous complex is defined. First section contains brief presentation of the problem and fixes the notation used in the paper. In the second section of the first chapter the the reader will find the construction of the ghost sector. The essential differences with conventional approach, which is suitable for anomaly-free systems are explained and justified. In particular: the identification of the space of physical states defined by mixed constraints in the differential space generated by the corresponding ghosts has to be neccesarily changed with respect to the conventional approach.  The corresponding normal orderings of operators have to be introduced. The subsection fixes also the neccessary correspondence bettween the languages used in the physical and mathematical literature.\\
The second chapter is devoted to the considerations on the anomalous complex with final result, which  identifies its cohomologies. As a the intermediate step, contained in the first subsection, the relative complex is introduced and its bigraded structure is analysed in details.  The K\"{a}hler
pairing of anomalous relative cochains and the corresponding Laplace operators
\footnote{One of them appeared as the universal operator generating Lagrange densities for the relativistic fields carrying arbitrary spin \cite{my}.} 
are introduced in the next subsections. The constructions, although parallell,  constitute essential generalization of those known in standard K\"{a}hler geometry \cite{Wells}. The realative cohomologies are identified with their harmonic representatives
\footnote{This result seems to be obvious at first glance but in fact it is far from being trivial} 
and vanishing theorem is proved. Finally the absolute cohomologies of anomalous complex are reconstructed. Some concluding comments are added in the last section. Technical and tedious calculations were extracted from the main text and are presented in two Appendixes.  \\ 
 \eject

\section{ BRST differential space}
This section is devoted to the construction and analysis of the differential space, which is
intrinsically related with anomalous Lie constraints or centrally extended Lie algebra action. The adjective anomalous is used in the physical literature. 
It simply denotes the centrally extended structure with central
term called the "anomaly" there. This section contains  mainly the canonical constructions related to Lie algebra cohomology, which are however modified and appropriately adopted. The first subsection is added here for the sake of completeness. The next ones are prepared  in order to stress the differences bettween standard (called nilpotent further on)  Lie algebra actions on the representation spaces and those with anomalies.
\subsection{Anomalous Lie constraints }
One starts with the representation of some real Lie
algebra $\mathfrak{g}$ (of compact type) on the complex space $V$, which is to be interpreted as the total space of states of constrained system. The
operators acing on $V$  do satisfy the
structural relations of  $\mathfrak{g}\,$:
    \begin{equation}
    \label{lie_rep}
    [\tilde{\mathcal{L}}_x,\tilde{\mathcal{L}}_y ] =
    \tilde{\mathcal{L}}_{[x,y]}\;; \;\;x,y \in \mathfrak{g}\,,
    \end{equation}
and may be interpreted as infinitesimal generators of symmetry transformations for example. 
Assume that for some or another reason the symmetry is broken and the constarints to be imosed as the conditions on the elements of $V$ are of the following form:
    \begin{equation}
    \label{cond}
    \tilde{\mathcal{L}}_x - <\chi,x>  \approx 0\;;\;\;x \in
    \mathfrak{g}\;\; {\rm for\; some}\;\; \chi \in\mathfrak{g}^\ast\,.
    \end{equation}
For non-abelian Lie algebra the above conditions are not contradictory but generally they admit, not acceptable, trivial solution cosisting of one element: $0$.  
This is clearly seen in terms of central extension or anomaly.\\ It
is convenient to introduce the following, shifted operators
    $\;\mathcal{L}_x := \tilde{\mathcal{L}}_x - <\chi,x>\;,\;\; x \in
    \mathfrak{g}\;$
    and then it is easy to see  the obstruction for the existence of non zero solutions explicitly:
    \begin{equation}
    \label{anomaly}
    [\,{\mathcal{L}}_x,{\mathcal{L}}_y\, ] =
    {\mathcal{L}}_{[\,x,y\,]} + <\chi, [\,x,y]>\;; \;\;x,y \in \mathfrak{g}.
    \end{equation}
The constraints of the above form are known as being of mixed type according to Dirac classification \cite{Dirac} and should be regarded with care in order to obtain a consistent anomaly free system.\\  
As it was already announced in the Introduction  it will be assumed that the Lie algebra $\mathfrak{g}$
is simple \cite{bourbaki} and $\chi$ is regular \cite{gs} in $\mathfrak{g}^\ast$ i.e it belongs to the interior of the appropriate Weyl chamber
\footnote{$\mathfrak{g}^\ast$ denotes algebraic dual of $\mathfrak{g}$. Under assumptions made above it can be identified with $\mathfrak{g}$ with the use of Killing form. In order to keep consistency with common  conventions \cite{bourbaki} this identification will never be exploited in this paper}. 
There exists Cartan subalgebra \cite{bourbaki} $\mathfrak{h}$ of $\mathfrak{g}$ such
that $\chi \in \mathfrak{h}^\ast\,$ in this case \cite{gs}. One may then always choose some basis $\{H_i\}_{i=1}^l$ in $\mathfrak{h}$ such that $\chi$ is
dominant i.e. $<\chi, H_i >\,\geq 0\,$ for all basis vectors. It will
be assumed that the basis of  $\mathfrak{h}$  with this property is chosen and
fixed.\\
With Cartan subalgebra being fixed one may perform the corresponding root decomposition\cite{bourbaki} of
$\mathfrak{g}$ and split the root system $R$ into disjoint subsets
of positive $R_+$ and negative roots $R_-$:
    \begin{equation}
    \label{root}
    \mathfrak{g} = \,\mathfrak{h}\oplus \bigoplus_{\alpha >0}
    \left(\mathbb{C}\tau_\alpha + \overline{\mathbb{C}}\tau_{-\alpha} \right)\;.
    \end{equation}
The structural relations of $\mathfrak{g}$ are then most conveniently encoded
in terms of Chevalley basis \cite{bourbaki}:
    \begin{eqnarray}
    [\,\tau_{\alpha}, \tau_{\beta}\,] &=& N_{\alpha \,\beta}\, \tau_{\alpha + \beta}
    - \delta_{\alpha\,,-\beta}\,H_{\alpha}\;,\;\; N_{\alpha \,\beta} = 0
    \;\;{if}\;\; \alpha + \beta \notin R
    \label{root_com_1}\\ \cr
    [\,H,\tau_{\alpha}\,]&=& \alpha (H)\,\tau_{\alpha}\,;\;\;
    H\in \mathfrak{h}\;.
    \label{root_com_2}
    \end{eqnarray}
    The opposite root vectors $\tau_{\alpha}$ and $\tau_{- \alpha}$
are chosen in such a way that the dual root $H_\alpha$ generated
by their commutator (\ref{root_com_1}) satisfies the canonical
normalization condition $\alpha (H_\alpha) = 2$.
	The structural relations
\footnote{The non-uniform notation will be used throughout this paper. The values of
$\vartheta \in \mathfrak{h}^*$ on $H \in \mathfrak{h}$ are generally denoted  by
 $<\vartheta,H >$ while, according to commonly used convention, the values of the roots $\alpha \in \mathfrak{h}^*$ are denoted by $\alpha (H)$.}
 of (\ref{anomaly}) in the Chevalley basis reads:
\begin{equation}
    [\,\mathcal{L}_{\alpha}, \mathcal{L}_{\beta}\,] = N_{\alpha \,\beta}\, \mathcal{L}_{\alpha + \beta}
    - \delta_{\alpha\;-\beta}\,\left(\,\mathcal{L}_{H_{\alpha}} + <\chi,H_{\alpha}>
    \right)\;,\;\;\;\;
    [\,\mathcal{L}_H,\mathcal{L}_{\alpha}\,]= \alpha(H)\,\mathcal{L}_{\alpha}\;,
    \label{action_com}
    \end{equation}
where $\mathcal{L}_{\alpha} := \mathcal{L}_{\tau_{\alpha}}$ for the sake of simplicity in the notation.\\
From the relations of (\ref{action_com}) it is clear that the maximal, anomaly free Lie subalgebra of constraints can be simply chosen as one of the complementary Borel subalgebras $\mathfrak{b}_\pm\, =\, \mathfrak{h} \bigoplus_{\alpha >0}
    \mathbb{C}\tau_{\pm \alpha}$ of $\mathfrak{g}\,$. This choice corresponds to well known Gupta-Bleuler idea applied in  quantum electrodynamics \cite{gb}.\\ The subspace of $V$ extracted by the constraints (\ref{cond}) is defined as the kernel of all elements of the distinguished and fixed subalgebra, say  $\mathfrak{b}_+\,$:
    \begin{equation}
    \label{kernel}
    V({\,\mathfrak{g},\chi }) = \left\{ \,\varphi\;;\;\; \mathcal{L}_\alpha \varphi = \mathcal{L}_H \varphi = 0\;,
    \;\;\alpha > 0 \,,\; H \in \mathfrak{h}\;\right\}\;,
    \end{equation}
according to natural generalization of \cite{gb}, which is even in much wider context, known in the mathematical literature as the polarization of Lie algebra \cite{gs}.\\  
 The elements of $V({\,\mathfrak{g},\chi })$ are to be immediately recognized as  highest weight vectors in $V$ of weight $\chi$ with respect to 
original Lie algebra action (\ref{lie_rep}) on $V$. It should be noted that complementary choice of $\mathfrak{b}_-$ leads to isomorphic, alhough not the same subspace of lowest weight vectors.\\
One more comment related with physical applications is in
order here. It it assumed that the space $V$ is equipped with
(positive) scalar product $(\,\cdot\,,\,\cdot\,)$.  \\
    If $A \rightarrow A^\ast$ is the corresponding conjugation, then the operators present
    in (\ref{action_com}) do satisfy the following conjugation
    rules:
 \begin{equation}
    \label{conjugations_V}
    \mathcal{L}_{\alpha}^{\ast} = \,- \,\mathcal{L}_{- \alpha}\;\;\;{\rm
    and}\;\;\; \mathcal{L}_H^{\ast} = \mathcal{L}_H \;;\;\; H \in \mathfrak{h}
    \;.
    \end{equation}
    These  rules  follow immediately from the properties of
    real structure constants in Chevalley basis \cite{bourbaki}: $N_{-\alpha\,-\beta} =
    N_{\alpha\,\beta}\,$.
    \subsection{The ghost sector}

The cohomological formultion of the the constrained systems (as well as the cohomological approach to Lie  algebras actions) is obtained by dressing the representation space with ghosts: $V \,\rightarrow\, \bigwedge
\mathfrak{g}^\ast \otimes V$. 
This space is equipped with nilpotent differential inherited from the exterior derivative \cite{GHV} of 
$\bigwedge\mathfrak{g}^\ast\,$ - this factor is commonly called the ghost sector in the physical literature.\\  The correspondence bettween abstract formulation in mathematical literature and the one in terms of ghost and anti-ghost creations adopted in the physical papers is established via Koszul formalism \cite{GHV}.
The Grassman algebra 
$\bigwedge
\mathfrak{g}^\ast\,$ is turned into irreducible representation module of  Clifford algebra defined by ghosts and anti-ghosts canonical anticommutation relations. The corresponding construction is briefly sketched here for the sake of completeness.\\
Let $c^i\,,\; i = 1,\ldots ,l = {\rm rank}(\mathfrak{g})$ denote the
operators of multiplication by the forms $\theta^i$ dual to the
Cartan subalgebra $\mathfrak{h}$ basis elements $H_i$. Let $b_i$
denote the dual  substitution operators corresponding to $H_i$. It is
clear that they satisfy the following anticommutation relations:
    \begin{equation}
    \label{clifford_0}
    \{\,c^i\,,\,b_j\,\} = \delta^i_j\;,\;\;\;\{\,c^i\,,\,c^j\,\}= 0 =
    \{\,b_i\,,\,b_j\,\}\;;\;\;\; i,j = 1\ldots l \;.
    \end{equation}
Similarly by  $b_{\alpha}\,,\; \alpha \in R\,$ one denotes the
substitution operators corresponding to the root vectors
$\tau_{\alpha}$. The operators $c^\alpha \,,\; \alpha \in R\,$ are
defined as the multiplication operators by the forms
$\theta^{-\alpha}$ dual to the root vectors
\footnote{This somewhat unusual convention is used in the physical literature,  especially in string theory}. 
 $\tau_{-\alpha}$.  Hence analogously to
(\ref{clifford_0}) one has:
    \begin{equation}
    \label{clifford_r}
    \{\,c^\alpha\,,\,b_\beta\,\} = \delta^\alpha_{-\beta}\;,\;\;\;\{\,c^\alpha\,,\,c^\beta\,\}= 0 =
    \{\,b_\alpha\,,\,b_\beta\,\}\;;\;\;\; \alpha,\beta \in R \;.
    \end{equation}
The relations (\ref{clifford_0}) together with (\ref{clifford_r})
define the structure of the Clifford algebra
corresponding to neutral quadratic form of $(d,d)$-signature,
where $d = \dim \mathfrak{g}\,$. In the physical literature the
generators $c^{(\cdot)}$ and $b_{(\cdot)}$ are called
ghost and respectively anti-ghost operators. In order to simplify the notation the the irreducible Clifford module of ghost anticomutation relations will be denoted by $\mathcal{C}$ instead of $\bigwedge\mathfrak{g}^\ast$.\\
 There is the representation of the Lie algebra $\mathfrak{g}$ (the extension of the coadjoint one) on $\mathcal{C}$. The corresponding Clifford algebra elements are given by Koszul formulae:
    \begin{eqnarray}
    {L}_{\alpha}^{\rm nil} &=& - \sum_{\beta \in R}\,N_{\alpha\,\beta}\,c^{-\beta}\,
    b_{\alpha + \beta} + \sum_{i=1}^l\,H_{\alpha}^i\,c^{\alpha}\,b_i
    + \sum_{i=1}^l\,\alpha (H_i)\,c^i\,b_{\alpha}\;;\;\;
    \alpha \in R \cr\cr
    {L}_i^{\rm nil} &=& - \sum_{\beta \in R}\,\beta(H_i)
    \,c^{\beta}\,b_{-\beta}\;;\;\; 1\leq i\leq l\;,
    \label{koszul}
    \end{eqnarray}
in the notation (\ref{root_com_1}-\ref{root_com_2}) of previous section. The numbers $H_{\alpha}^i$ are the coordinates of $H_{\alpha}$ in some basis $\{H_i\}$ of Cartan subalgebra
\footnote{The superscript $(\cdot)^{\rm nil}$ has been introduceded in order to distinguish bettween the anomaly free (nilpotent) constuction and the anomalous one which will be presented in the core of  this paper.}.
\\
The corresponding (exterior) differential is given by the following Clifford algebra element \cite{GHV}:
\begin{equation}
    \label{koszul_diff}
    {d}^{\;\rm nil} = \frac{1}{2}\,\sum_{\alpha \in R}\, c^{-\alpha}\,{L}_{\alpha}^{\rm nil}
    + \frac{1}{2}\,\sum_{i=1}^l\, c^i\,{L}_i^{\rm nil}\;\;\;{\rm and\;it\;is\;nilpotent}\;\;\;
    (\,{d}^{\;\rm nil}\,)^{\,2} = 0\;.
\end{equation}
One more step (neglecting the details which are to be presented within the analysis the anomalous case) is neded to finish the construction in the absence of anomaly: $\chi = 0$. The original representation space $V$ is dressed by ghosts: 
\begin{equation}\label{standard_complex}
V \,\rightarrow\,  \mathcal{C}(V)\,:=\,\mathcal{C}\,\otimes V\, = \,\bigoplus_{r}\,\mathcal{C}^r(V)\;, \;\;\; {\rm where }\;\;\;
\mathcal{C}^r(V) = {\bigwedge}^r
\mathfrak{g}^\ast \otimes V\,,
\end{equation}
and it is equipped with canonical differential:
\begin{equation}
    \label{total_diff_nil}
    {D}^{\rm nil} = \sum_{\alpha \in R}\, c^{-\alpha}\,\otimes\,\mathcal{L}_{\alpha}\,
+ \sum_{i = 1}^{l}\,c^i\,\otimes \mathcal{L}_{i}\,\;+\;{d}^{\rm nil}\,\otimes\,1\;,
    \end{equation}
where $\mathcal{L}_{(\cdot)}$ are that of (\ref{cond}) with $\chi = 0$ and ${d}^{\rm nil}$ is that of (\ref{koszul_diff}).\\ The space $V$ is identified with $\mathcal{C}^0(V)$ of (\ref{standard_complex}) and the invariant elements of $V$ defined by anomaly free constraints are determined by single equation: ${D}^{\rm nil}\,\varphi = 0$. It is also clear that they constitute the content of cohomology space at degree zero
\footnote{According to well known results \cite{fuks}, the higher classes are non zero even for simple Lie algebras. }.
\\\\
The way of proceeding has to be modified in the anomalous case ($\chi \neq 0$). 
In order to obtain an analogous description of the Borel
subalgebra invariant subspace (\ref{kernel}) of $V$  within the same 
ghost dressed space $\bigwedge \mathfrak{g}^\ast \otimes V$ of left invariant
forms with values in $V$ one cannot identify the space $V$ as the one consisting of the elements of degree zero in (\ref{standard_complex}). The equation 
$D^{\rm nil}\,\varphi = 0$ would imply too many conditions leading to unsatisfactory solution $\varphi = 0$ instead of  (\ref{kernel}).
\\ 
A closer look at the expression (\ref{total_diff_nil}) for differential suggests that the space $V$ should be identified witin $ \mathcal{C}(V)$ with the tensor product $\omega \otimes V$, where $\omega$ is such that it is  anihilated by all terms corresponding to negative roots i.e. by all terms of  $\, \sum_{\alpha > 0}\, c^{\alpha}\,\otimes\,\mathcal{L}_{-\alpha}$. At the same time the remaining terms (with $d^{\rm nil} \otimes 1\,$ to be corrected, is neglected for the moment) of  (\ref{total_diff_nil}) should reproduce the conditions extracting the elements of the space $ V({\,\mathfrak{g},\chi })$ of (\ref{kernel}) under action on  $\omega \otimes V$. \\
\\
The following
\begin{definition}[Ghost vacuum]\ \\
\\
A non zero element $\omega \in \mathcal{C}$ satisfying
    \begin{equation}
	\label{ghost_vac}
    c^{\alpha}\,\omega = b_{\alpha}\,\omega = 0 \;;\;\;\alpha > 0
    \;\;\;\;{\rm and}\;\;\;\; b_i\,\omega = 0\;;\;\; 1\leq i \leq l\,,
    \end{equation}
    is called the ghost vacuum.
    \end{definition} 
is natural and unique in the light of the above arguments.  The vacuum is fixed by (\ref{ghost_vac}) up to scalar factor and can be chosen to be the top form over all negative roots subspace:
    \begin{equation}
    \label{vacuum}
    \omega = \theta^{-\alpha_1}\wedge\ldots\wedge\theta^{-\alpha_m}\;,\;\;\;\alpha_i \in R_+\;,
    \end{equation}
where some order of roots is taken into account and fixed once for all.
It is also clear that the space $\mathcal{C}$ is generated out of
(\ref{vacuum}) by the action of $\,c^{-\alpha}\,,\,b_{-\alpha}\,$
and $c^i\,$ ghost and anti-ghost creation operators. \\
There is however price one has to pay for the choices made obove.  It was already implicitely noted that there might be an obstacle in the identification of 
$\mathfrak{b}_+ $ - invariant elements of $V$ with $D^{\rm nil}\,$-closed elements of $\omega \otimes V$. In fact, neither $\omega$ is closed: 
$d^{\rm nil}\omega \neq 0$ nor it is invariant with respect to Cartan sualgebra elements of (\ref{koszul}): $L_i^{\rm nil}\omega \neq 0$.
\\
In order to correct this property one introduces the normal ordering rule for Clifford algebra generators (ghost operators)
    \begin{equation}
    \label{normal}
    :c^{\alpha}b_{\beta}: \;\;= \left\{
    \begin{array}{ll}
   \; \;\;c^{\alpha}b_{\beta}\;;& \beta > 0 \cr
    - b_{\beta}c^{\alpha}\;;& \beta < 0
    \end{array}
    \right.\;\;,\;\;\;\;\;\; :c^i b_j:\;\; = \frac{1}{2}(c^i b_j - b_j c^i)
    \;,
    \end{equation}
and consequently all the Clifford elements of (\ref{koszul}) are replaced by their normally ordered counterparts:
    \begin{equation}
    \label{koszul_n}
    L_{\alpha} = \;:{L}_{\alpha}^{\rm nil}:\;\;;\;\;\; \alpha \in R
    \;,\;\;\; L_i = \;:{L}_i^{\rm nil}:\;\;;\;\;\;1\leq i\leq l\;.
    \end{equation}
From the expressions (\ref{koszul}) it immediately follows that
the normal ordering does not affect  the operators corresponding to
root vectors: $\,L_{\alpha} \equiv {L}_{\alpha}^{\rm nil}$. For the
Cartan subalgebra elements the normal ordering is non trivial and one obtains: 
    \begin{equation}
    \label{cartan_n}
    L_{i} = -\sum_{\beta > 0}\beta (H_i)c^{-\beta}b_{\beta} -
    \sum_{\beta > 0}\beta (H_i) b_{-\beta}c^{\beta}\;;\;\;\;\;1\leq i\leq l\;.
    \end{equation}
 The ghost vacuum satisfies now the desired equations $L_i\, \omega = 0\,$ but (\ref{cartan_n}) differ from those of (\ref{koszul}), which results in 
breaking the original commutation relations (\ref{lie_rep})
    and the algebra defined by (\ref{koszul_n}) gets centrally extended, i.e. it becomes anomalous in the physical language.\\
The structure defined by normally ordered operators can be
    displayed and summarized in the following
    \begin{lemma}\
    \begin{enumerate}
    \item
    $\;\;L_i = {L}_i^{\rm nil} - 2 <\varrho,H_i>\;,\;\;$ where
    $\;\;\varrho =\frac{1}{2}\sum\limits_{\alpha > 0}\alpha\;\;\;$
    is the lowest dominant weight.
    \item
    The normally ordered operators satisfy the following
\footnote{In the case of of infinite-dimensional Lie algebra, with Virasoro algebra being the prominent example, the formulae {\it 1. of Lemma} does not make much sense as
it gives  divergent series and it needs some regularization (by $\zeta$- function for example).} 
    structural relations:
    \begin{equation}
    \label{ghost_centr}
    [\,L_i,L_{\alpha}\,] = \alpha (H_i)\,L_{\alpha}\;\,,\;\;\;\;
    [\,L_{\alpha}, L_{\beta}\,] = N_{\alpha\,\beta}\,L_{\alpha +\beta}
    - \,\delta_{\alpha\,-\beta}\,(\,L_{H_{\alpha}}
    + 2<\varrho, H_{\alpha}>\,)\;.
    \end{equation}
    \end{enumerate}
    \end{lemma}
{\it Proof:}\\
{\it 1}. The relation can be obtained  from (\ref{cartan_n}) by
the use of anti-commutation relations (\ref{clifford_r}):
    $\;L_i = - \sum_{\beta > 0}\beta (H_i)
    (c^{-\beta}b_{\beta} -
    c^{\beta}b_{-\beta}
    + \{c^{\beta},b_{-\beta}\}) = {L}_i^{\rm nil}
    - (\sum_{\beta > 0}\beta )(H_i) =
    = {L}_i^{\rm nil} - 2 <\varrho,H_i> \;$.\\
{\it 2}. The only relations which are changed by normal ordering
with respect to those of (\ref{root_com_1}) and (\ref{root_com_2})
are that containing the Cartan subalgebra elements on the right
hand side i.e. those of opposite root vectors:
    $ [\,L_{\alpha},L_{-\alpha}\,] = [\,{L}_{\alpha}^{\rm nil},{L}_{-\alpha}^{\rm nil}\,] =
    - {L}_{{H}_{\alpha}}^{\rm nil} = - (L_{H_{\alpha}} + 2 <\varrho,H_\alpha>)
    \,$ ($\alpha > 0$). The last equality is written due to relation of {\it{1}}.
    $\;\;\blacksquare$\\ 
\\
   The differential (\ref{koszul_diff}) does not kill the vacuum element $\omega$ of (\ref{vacuum}). In order
to introduce a differential $d$ with the property $d\,\omega = 0$
one has to use the normal ordering again. The appropriate Clifford
algebra element is given in the following
    \begin{definition}[Anomalous ghost differential]\ \\
\\
    The operator
    \begin{equation}
    \label{koszul_diff_n}
    d :=  \,\; :\!{d}^{\rm nil}\!:\,\;= \,\frac{1}{2}\,\sum_{\alpha > 0}\,
    c^{-\alpha}\,L_{\alpha} + \frac{1}{2}\,\sum_{\alpha > 0}\,
    L_{-\alpha}\,c^{\alpha} +
    \frac{1}{2}\,\sum_{i=1}^l\, c^i\,{L}_i\;\;\;\;\;\;\;\;\;\;\;\;
    \end{equation}
    is called the anomalous ghost differential.
    \end{definition}
From the above definition it immediately follows that:
    $d\,\omega = 0\;$.\\
In order to find the relation between nilpotent differential (\ref{koszul_diff}) and the
anomalous one it is convenient to introduce a nilpotent
Clifford algebra element corresponding to the dominant weight  
$\varrho$, namely: ${\varrho}_{\rm op} =
\sum_{i=1}^l\,<\varrho,H_i>\,c^i\,$.\\
\begin{lemma}
\begin{equation}
    \label{d_d_hat_relation}
    d = {d}^{\rm nil} - 2\,{\varrho}_{\rm op}
    \end{equation}
\end{lemma}
{\it Proof:}\\
The expression for $d$ in terms of ${d}^{\rm nil}$ and ${\varrho}_{\rm op}$ can be obtained by
reordering  (\ref{koszul_diff_n}) to that of
(\ref{koszul_diff}) and with the help of the relation of
    {\it Lemma 1.1}:
    $
    \;2\,d = \,\sum_{\alpha > 0}\,
    c^{-\alpha}\,L_{\alpha} + \,\sum_{\alpha > 0}\,
    ( \,[\,L_{-\alpha},\,c^{\alpha}\,] + c^{\alpha}\,L_{-\alpha})
    + \,\sum_{i=1}^l\, c^i\,
    (\,{L}_i^{\rm nil} - 2<\varrho,H_i>) = 2\,{d}^{\rm nil} \,- \,2\,{\varrho}_{\rm op}
    \,+ \,\sum_{\alpha > 0}\,[\,L_{-\alpha},\,c^{\alpha}\,]\;$.
 Using (\ref{koszul}) one immediately calculates that
    $[\,L_{-\alpha}, c^{\alpha}\,] = - \sum_{i=1}^l c^i\, <\alpha,
    H_i>\,$, which after summation over all positive roots gives additional contribution of $- \,2\,{\varrho}_{\rm op}\,
    $.  $\;\;\blacksquare$\\ 
 The above two lemmas lead one to the conclusion that   the ghost differential $d$ is not nilpotent and one has instead:
\begin{proposition}
\begin{equation}
\label{curvature}
    d^{\,2} \,=\,
     - 2\,\sum_{\alpha > 0}\,c^{-\alpha}\,c^{\alpha}\,
    <\varrho\,,H_{\alpha}>\;.
\end{equation}
\end{proposition}
    {\it Proof} : Taking into account that ${d}^{\rm nil}$ as well as ${\varrho}_{\rm op}$
    are nilpotent one has to calculate explicitely the expression for
    $
    d^{\,2} = - 2\, (\,{\varrho}_{\rm op}\,{d}^{\rm nil} + {d}^{\rm nil}\,{\varrho}_{\rm
    op}\,)\;
    $ only.
    According to the definition of ${\varrho}_{\rm op}$ it is enough to find
    $
    2\,\{\,c^i\,,{d}^{\rm nil}\,\} = - \sum_{\alpha > 0}\,c^{-\alpha}\,[\,c^i, L_{\alpha}\,]
    - \sum_{\alpha > 0}\,c^{\alpha}\,[\,c^i,L_{-\alpha}\,] =  2\,
    \sum_{\alpha > 0}\,c^{-\alpha}\,c^{\alpha}\,H_{\alpha}^i\;
    $.
    The last equality is obtained from (\ref{koszul}) by direct
    calculation with the help of  structural relations (\ref{clifford_0}).
    Taking into account the definition of $H_{\alpha}^i$
    coefficients and the definition of $\varrho$ one obtains the
    thesis. $\;\;\blacksquare$\\ 
All the constructions presented above can be reinterpreted in more geometrical language used in mathematical literature. The element $2\,{\varrho}_{\rm op}$ can be thaught of as the connection form in appropriate vector bundle. The normally ordered operator $d$ is then interpreted as the covariant differential
and the right hand side of (\ref{curvature}) is simply the corresponding curvature. However this point of view will not be pursued. The name "crvature" for (\ref{curvature}) will be used from time to time in this paper.\\
Finally one may introduce new grading of the ghost sector $\mathcal{C}$. Although for the considerations of this paper, in contrast to infinite systems of constraints, it does not play any essential role it will be adopted in this article for the sake of agreement with conventions of physical literature. The new (non - positive and not necessarily integral) grading is obtained by splitting $\mathcal{C}$ into direct sum of eigenspaces of the ghost number operator:
\begin{equation}
\label{ghost_number}
    {\rm gh}_{\rm tot}\, = \,\sum_{\alpha > 0}\, c^{-\alpha}\,b_{\alpha} -
    \sum_{\alpha > 0}\,b_{-\alpha}\,c^{\alpha} +
    \frac{1}{2}\,\sum_{i=1}^{l}\,(\,c^i\,b_i - b_i\,c^i\,)\;,
\end{equation}
which is nothing else than normally ordered total degree operator
\footnote{The subscript $ ~_{\rm tot}$ is added in order to make the difference of (\ref{ghost_number}) with another grading operator used further on in this paper.}.
 Nevertheless the  module $\mathcal{C}$ can be
equipped with integral, although non positive, grading such that
$d$ is an odd derivation operator of degree $+1$. 
The
ghost vacuum $\omega$ of (\ref{vacuum}) is the eigenvector of
${\rm gh}_{\rm tot}$ with eigenvalue $- \frac{1}{2} l$. The ghost and
anti-ghost creation/anihilation operators are of ghost weights
$+1$ and $-1$ respectively. This property follows immediately from
the following commutation relations:
    \begin{equation}
    \label{ghost_weigts}
    [\,{\rm gh}_{\rm tot}\,, c^{(\cdot)}\,] = c^{(\cdot)}\;,\;\;\;[\,{\rm gh}_{\rm tot}\,, b_{(\cdot)}\,] =
   -  b_{(\cdot)}\;.
    \end{equation}
Consequently, the space $\mathcal{C}$ splits into the direct sum of
eigensubspaces $\mathcal{C}^r$ of ghost number operator
(\ref{ghost_number}):
    \begin{equation}
    \label{ghost_split}
    \mathcal{C}\,=\,\bigoplus_{r = -m - \frac{1}{2}l}^{r= m +\frac{1}{2}l}\,\mathcal{C}^r\;,\;\;
    \;\;\;{\rm where}\;\;\;l = {\rm rank}\,\mathfrak{g}\;\;{\rm and}\;\; 2m + l = \dim \mathfrak{g}\;,
    \end{equation}
corresponding to eigenvalues $r = -m - \frac{1}{2}l + i\;,\;\;
1\leq i \leq \dim \mathfrak{g}\,$.\\
Using the definition (\ref{koszul_diff_n}) and the properties
(\ref{ghost_weigts}) of ghost operators one immediately may see that
    $[\,{\rm gh}_{\rm tot},{d}\,]\, =\, d$
  i.e.  the differential $d$ raises the ghost number
 by $+1\;$. This property is transparently illustrated
in the following diagram:
    $$
    \mathcal{C}^{-m-\frac{1}{2}l} \stackrel{d}{\rightarrow}\cdots
    \mathcal{C}^{r}\stackrel{d}{\rightarrow}\mathcal{C}^{r+1}
    \cdots \stackrel{d}{\rightarrow}\mathcal{C}^{m+\frac{1}{2}l}
    \stackrel{d}{\rightarrow}0\;.
    $$
The space $\mathcal{C}$ equipped with differential $d$ of
(\ref{koszul_diff_n}) is graded differential space although, due
to (\ref{curvature}) {\bf it is not }a complex.\\
\\
One more remark is in order at the end of this subsection.
It is possible to introduce the neutral pairing on the differential
space $\mathcal{C}$. The formal conjugation rules (consistent with those of (\ref{conjugations_V})) of ghost and anti-ghost operators: 
\begin{equation}
\label{conj_rules}
    {c^{\alpha}}^{\ast}= -\, c^{- \alpha}\;,\; b_{\alpha}^{\ast} = -
   \, b_{-\alpha}\,\;,\;\; \alpha \in R \;,\;\;\;\; {c^i}^{\ast} = c^i\;,\; b_i^{\ast} = b_i\;,\;\; i = 1\ldots l\;,
    \end{equation}
supplemented with  the normalization condition of the ghost vacuum: 
    \begin{equation}
    \label{ghost_vacuum_norm}
    (\,\omega\,,\,\upsilon (\mathfrak{h})\,\omega\,) = 1\;\;\;{\rm where}\;\;\; 
	\upsilon (\mathfrak{h}) := c^1\ldots c^l\;.
    \end{equation}
define (with the help of 
relations of (\ref{clifford_r})) the unique, non-degenerate, hermitean scalar product on $\mathcal{C}$.\\
It is easy to check that the Lie algebra generators (\ref{koszul})
as well as their normally ordered counterparts satisfy the same
conjugation relations as the operators acting in $V$ i.e. those of
(\ref{conjugations_V}). The differential $d$ 
(\ref{koszul_diff_n}) is consequently self-adjoint: $d^{\ast} = d$.
\subsection{$V$-differential space}

It is now possible to come back to the original $\mathfrak{g}$ - module $V$ after preparations made in the former sobsection. As it was already mentioned the space $V$ gets dressed with ghosts:
    \begin{equation}
    \label{ghost_split_v}
    \mathcal{C}(V)\,:=\,\mathcal{C}\,\otimes V\, = \,\bigoplus_{r}\,\mathcal{C}^r(V)\;,
    \end{equation}
where the corresponding
decomposition is naturally inherited from that of
(\ref{ghost_split}). It will be assumed that $\mathcal{C}(V)$ is
equipped with the natural pairing induced on the tensor product by
the pairings on the factors. It will be denoted by the round bracket too.
Consequently the conjugation of operators acting on $\mathcal{C}(V)$ will be denoted by $(\cdot)^{\ast}$. \\
The differential in the space $\mathcal{C}(V)$ of (\ref{ghost_split_v}) is defined in the standard way:
    \begin{equation}
    \label{total_diff}
    D = \sum_{\alpha \in R}\, c^{-\alpha}\,\otimes\,\mathcal{L}_{\alpha}\,
+ \sum_{i = 1}^{l}\,c^i\,\otimes \mathcal{L}_{i}\,\;+\; d\,\otimes\,1\;,
    \end{equation}
where $\mathcal{L}_{(\cdot)}$ are that of (\ref{cond}) and $d$ is normally ordered this time i.e. that of (\ref{koszul_diff_n}). From the above definition it follows that the space $\mathcal{C}(V)$ equipped with the
differential $D$ is graded differential space:
$D\,:\,\mathcal{C}^r(V)\,\rightarrow\,\mathcal{C}^{r+1}(V)\,$.  From here and in the sequel this space will be called {\bf totall differential space} of the problem. It
is worth to note that the spaces $\mathcal{C}^r(V)$ and
$\mathcal{C}^{-r}(V)$ are mutually dual with respect to the scalar
product on $\mathcal{C}(V)$. This property follows immediately
from the fact that the ghost number operator 
(\ref{ghost_number}) does satisfy $ {\rm gh}^{\ast}_{\rm tot}\, =
- \, {\rm gh}_{\rm tot}\,$. It is also worth to note that the
differential (\ref{total_diff}) is symmetric $D^{\ast} = D$ .
    \\
The the basis elements of the Lie algebra $\mathfrak{g}$ acting on the total (ghost dressed) space are obtained from (\ref{total_diff}) by the standard formulae \cite{GHV}:  
\begin{equation}
    \label{total_l}
    L_{(\cdot)}^{\rm{ tot}} \,:=\,\{\,b_{(\cdot)},D\,\} =
   1\otimes \mathcal{L}_{(\cdot)}\,
   +\,L_{(\cdot)}\,\otimes\,1\;,
    \end{equation}
where $L_{(\cdot)}$ are that of (\ref{koszul_n}). From
(\ref{action_com}) and (\ref{ghost_centr}) one immediately obtains
that the Lie algebra defined by (\ref{total_l}) is centrally
extended with the anomaly containing contibution from both: matter factor and ghost sector:
    \begin{equation}
    \label{total_anomaly}
    [\,L_{\alpha}^{\rm {tot}}, L_{\beta}^{\rm {tot}}\,] = N_{\alpha\,\beta}\,L_{\alpha +\beta}^{\rm tot}
    - \,\delta_{\alpha\,,-\beta}\,(\,L_{H_{\alpha}}^{\rm {tot}} +
    <\chi + 2\varrho \,, H_{\alpha}>\,)\;.
    \end{equation}
The form of the anomaly might suggest that under favourable conditions they can cancel each other as it happens in sting theory in critical dimensions. Nothing like that can happen here as the anomalies are cohomologically trivial and  since they are located inside the same Weyl chamber both are positive \cite{gs}. In string theory the anomaly associated with coadjoint action in the ghost sector is non trivial and negative ($= -26$ or $-10$ with respect to appropriate normalization).\\
In order to simplify the notation the mark $\otimes$ of tensor product will be supressed in the sequel.\\
The properties of the operator  (\ref{total_diff}) are summarized
in the following
    \begin{proposition}\ \\
\\
    The differential $D$ of (\ref{total_diff}) is neither
    nilpotent nor it is invariant with respect to (\ref{total_l}):
    \begin{equation}
    \label{total_curvature}
    D^{\,2} \,= \,
    -  \sum_{\alpha > 0}\,c^{-\alpha}\,c^{\alpha}\,
    <\chi +2\varrho\,,H_{\alpha}>\;,\;\;{\it and}\;\;\;
    [\,L_{\alpha}^\tot\,,\,D\, ] \, = \,-
     c^{\alpha}\,
    <\chi + 2\varrho\,,H_{\alpha}>\;.
    \end{equation}
    \end{proposition}
    {\it Proof :}
    The formulae for the curvature can be obtained by finding its
    components: $D^2 (x,y) = \{ \,b_y \,, [\,b_x\,,\,D^2\,]\,\}$ with 
$x,y \in \mathfrak{g}$.
    Using definition (\ref{total_l}) and graded Leibnitz identity
    one first of all finds: $[\,b_x\,,\,D^2\,] = [\,L_x^\tot\,,\,D\,]$.
    Then by graded Jacobi identity in the Clifford algebra one
    obtains:
    $\{ \,b_y \,, [\,b_x\,,\,D^2\,]\,\} = [\,L_x^\tot\,,\,L_y^\tot\,] -
    \{\,[\,L_x^\tot\,,\,b_y\,]\,,\,D\,\}
    $.
Since $[\,L_x^\tot\,,\,b_y\,] = b_{[\,x\,,\,y\,]}$, the components
    of the curvature are equal to anomaly present in
    (\ref{total_anomaly}). The second formulae  easily follows from
    just obtained expression for $D^2$ and from already used identity
    $[\,b_x\,,\,D^2\,] = [\,L_x^\tot\,,\,D\,]$. $\;\;\;\blacksquare$\\
\\
One should note that the differential is invariant under Cartan
subalgebra i.e. $[\,L_H^\tot\,,\,D\,]= 0\,$ for any $H \in \mathfrak{h}\,$.\\
For later convenience it is worth to introduce the
notation related to the anomaly coefficients, namely:
    \begin{equation}
    \label{coefficients}
    r_{\alpha} := {\rm sign }(\alpha)<\chi +2\varrho\,,H_{\alpha}>\;,\;\;\alpha
    \in R \;\;\;\;{\rm so\; that} \;\;\;\; D^{\,2} = - \sum_{\alpha >
    0}\,r_{\alpha}\,c^{-\alpha}\,c^{\alpha}\;\;.
    \end{equation}
It also worth to  notice that the anomaly (curvature) coefficients do satisfy the
important identity which follows from the cocycle property
(Bianchi - identity) of $D^2$.
\begin{lemma}\ \\
\\
For any $\alpha\,,\,\beta\,,\,\gamma \in R\,$:
    \begin{equation}
    \label{cocycle}
   {\rm sign }(\alpha + \beta) r_{\alpha + \beta}\,N_{\alpha\,\beta}\, = 
{\rm sign }(\alpha + \gamma )\,r_{\alpha + \gamma}\,N_{\alpha\,\gamma}
    \,- \,{\rm sign }(\beta + \gamma) r_{\beta + \gamma}\,N_{\beta\,\gamma}\;,
    \end{equation}
where $N_{\cdot\,\cdot}$ are the structure constants introduced in
(\ref{root_com_1}).
\end{lemma}
{\it Proof:}  The Jacobi identity implies that for any $\lambda \in
\mathfrak{h}^\ast$ one has the following equation:
    $
    <\lambda\,,[\,\tau_{\alpha}\,,\,[\,\tau_{\beta}\,,\,\tau_{\gamma}\,]\,]>\,=\,
    -\,<\lambda\,,[\,\tau_{\gamma}\,,\,[\,\tau_{\alpha}\,,\,\tau_{\beta}\,]\,]>\,+\,
    <\lambda\,,\,[\,\tau_{\beta}\,,\,[\,\tau_{\alpha}\,,\,\tau_{\gamma}\,]\,]>\,,
    $
    which according to (\ref{root_com_1}) immediately implies the
    desired identity for curvature coefficients corresponding to
    $\lambda = \chi + 2 \varrho\,$. $\;\blacksquare$\\
\\
    The results above say that the space $\mathcal{C}(V)$
equipped with the differential $D$ {\bf is not} a complex again. Nevertheless
it has the structure of graded differential space and again one
may draw the diagram:
    $$
    \mathcal{C}^{-m-\frac{1}{2}l}(V) \stackrel{D}{\rightarrow}\cdots
    \mathcal{C}^{r}(V)\stackrel{D}{\rightarrow}\mathcal{C}^{r+1}(V)
    \cdots
    \stackrel{D}{\rightarrow}\mathcal{C}^{m+\frac{1}{2}l}(V)
    \stackrel{D}{\rightarrow}0\;.
    $$
The next section will be devoted to the construction of the appropraite complex within the above differential space.
    \section{Anomalous complex}
Despite of the evident (\ref{total_curvature}) obstructions  the canonical complex associated with the anomalous problem will be defined and described in this section. In fact there are at least two possibilities coming into mind. The first one, to be recognized  as most
obvious and natural, consists in replacing the total differential space (\ref{ghost_split_v}) by the maximal subspace on which the differential $D$ is nilpotent: the kernel of the curvature (\ref{total_curvature}). 
The second possibility is a bit more sophisticated and consists of the choice  of  the maximal
subspace of (\ref{ghost_split_v}) such that the differential is
invariant with respect to Borel subalgebra $\mathfrak{b}_+$ of $\mathfrak{g}$. The considerations of this paper will be concentrated exclusively on this very former complex. The remarks in the {\it Conclusions} show that the second choice leads to its subcomplex.\\
The above comments are concluded in the form of the following
    \begin{definition}[Anomalous complex]\ \\
\\
    The differential space
    \begin{equation}
    \label{anomalous_complex}
    (\,\mathcal{A}\,,\,D\mid_{\mathcal{A}}\,)\;\;\;{\rm where}\;\;\;\mathcal{A} =
    \ker D^2\;,
    \end{equation}
    is to be said the anomalous complex.
    \end{definition}
    For the sake of simplicity of the notation the mark of restriction
    $\,.\mid$ of the domain except of the cases where it is necessary will be suppressed in the sequel. The elements of $\mathcal{A}$ will be sometimes called cochains according to mathematical nomenclature.
\\
    In order to obtain more information on the anomalous complex defined
    in (\ref{anomalous_complex})  one introduces the following operators corresponding to the anomaly (\ref{total_anomaly}), which in slightly different
\footnote{ Very close context in fact. }
context of complex differential geometry were in simpler form known from  long time ago \cite{Wells}:
    \begin{eqnarray}
    \label{sl_2}
    J^+ \,&:=& - D^2 =  \sum_{\alpha>0}\,r_\alpha\, c^{-\alpha}\,c^{\alpha}\;,\;\;
    J^- \,:= \sum_{\alpha>0}\,\frac{1}{r_\alpha }\,
    b_{-\alpha}\,b_{\alpha}\;,\;\;\\
    \nonumber
    {\rm gh}_{\rm rel}\,&:=& {\rm gh}_{\rm tot} - \frac{1}{2}\,\sum_{i=1}^{l}\,(\,c^i\,b_i -
    b_i\,c^i\,)\, = \,\sum_{\alpha > 0}\,( c^{-\alpha}\,b_{\alpha}\, -
    \,b_{-\alpha}\,c^{\alpha}\,)\;,
    \end{eqnarray}
    where ${\rm gh_{\rm tot}}$ is that of (\ref{ghost_number}) and $r_\alpha$ are the anomaly coefficients defined in (\ref{coefficients}). 
\footnote{ Somewhat, at the moment exotic notation $ {\rm gh}_{\rm rel}$ is justified in the next subsection, where the corresponding Clifford element serves as natural grading operator of so called relative complex.}.   
By direct calculation with the help of commutation rules (\ref{clifford_r}) on obtains the following
    \begin{lemma}\ \\
\\
    The operators defined in (\ref{sl_2}) satisfy the structural relations
    of $sl(2)$ Lie algebra:
    \begin{equation}
    \label{sl_2_com}
    [\,J^+\,,\,J^-\,] = {\rm gh}_{\rm rel}\;\,,\;\;\;
    [\,{\rm gh}_{\rm rel}\,,\, J^\pm\,] = \pm 2\,J^\pm\;\;\;.\;\;\;\;\;\;\;\;\;\;\;\;\;\;\;\blacksquare\;
    \end{equation}
    \end{lemma}
 The operator $J^+$ introduced in (\ref{sl_2}) is (up to sign) nothing else than the curvature (\ref{total_curvature}). As it was the case in K\"{a}hler geometry \cite{Wells}, this algebra will be, with essential modifications,  intensively exploited in the computations of the cohomologies of the anomalous complex.\\   
At the moment one may obtain some information on the structure of the anomalous complex based on simple facts on the representations of $sl(2)$ Lie algebra. First of all one should notice that the operator
	${\rm gh}_{\rm rel}$ of (\ref{sl_2_com}) is  diagonal on the full differential space 
	$\mathcal{C}(V)$ and its spectrum ranges from $-m$ to $m$ step $1$, where 
	$2m = \dim \mathfrak{g}/\mathfrak{h}\,$. From the definition (\ref{anomalous_complex}) it 
	follows that the space $\mathcal{A}$ consists of highest weight vectors with respect to 
	$sl(2)$ Lie algebra of (\ref{sl_2_com}). It is well known fact \cite{bourbaki} that all highest weight vectors are of non-negative weights. Hence 
	$\mathcal{A} = \bigoplus_{j\geq 0}\,{\mathcal{C}}_+(j)\otimes V\,$, where 
   ${\mathcal{C}}_+(j)$ contains all highest weight vectors of the ghost differential space
	(\ref{ghost_split}) of weight $j\,$. From the definition (\ref{ghost_vac}) it folows that
	 the ghost 
	vacuum $\omega$ ($j=0$) as well as all the elements generated out of the vacuum by 
	$c^i$ and $c^{-\alpha}$-ghost creation operators i.e. those from 
	$\bigwedge \mathfrak{b}_+^*\,\omega\,$ are of highest weights. In particular the elements 
	of the space
\footnote{$\mathfrak{n}_+^*$ denotes the dual of nilpotent Lie algebra generated by all positive root vectors.}
	$\bigwedge^s \mathfrak{h}^* \bigwedge^j\mathfrak{n}_+^* \omega$ are all of weight $j$, 
	while their total ghost degree (\ref{ghost_number}) equals to 
	$r = -\frac{1}{2}l + s + j\,$. The elements of this form do not exhaust the
	 space of highest weight vectors as there are for example the vectors generated by 
	ghost - anti-ghost clusters  of weight zero:   
    \begin{equation}
	\label{clusters}
	G_{\alpha} := c^{-\alpha}b_{-\alpha}\;\;\;{\rm and}\;\;\;
	G_{\alpha\;\beta} := \frac{1}{r_{\alpha}}\,b_{-\alpha}c^{-\beta} + 
	\frac{1}{r_{\beta}}\,b_{-\beta}c^{-\alpha}\;,\;\; \alpha \neq \beta > 0\;.
	\end{equation}
	It is left as an open question whether the above operators together with ghost operators generate the whole space of the anomalous complex out of the vacuum. The details on the corresponding  $sl(2)$ - modules can be found in \cite{bourbaki}. \\
	Nevertheless from the considerations above it follows that the spectrum of the admissible ghost numbers in the anomalous complex is asymmetric and at negative degrees it terminates at $-\frac{1}{2}l$ - the ghost number of the vacuum:
	\begin{equation}
	\label{asymm}
	\mathcal{A} = \bigoplus_{r = -\frac{1}{2}l}^{r= m +\frac{1}{2}l} \mathcal{A}^r\;.
	\end{equation}
For this reason the scalar product introduced in $\mathcal{C}(V)$ becomes highly 
degenerate when restricted to the anomalous complex.  Further on a non degenerate and positive pairing determined by anomaly will be introduced on the distinguished, so called relative, subspace of  $\mathcal{A}$.
\\
For the sake of completeness the definition of anomalous cohomology spaces has to be given. In the light of (\ref{anomalous_complex}) it cannot be different than the following one:
    \begin{definition}[Anomalous cohomologies]\ \\ 
\\
    The quotient spaces
    \begin{equation}
    \label{anom_coho}
    H^r  = \frac{Z^r}{B^r}\;,\;\;\;\;\;\;Z^r = \ker D\! \mid_{\mathcal{A}^r}\;,\;\;B^r = {\rm im}\;
    D\! \mid_ {\mathcal{A}^{r-1}}
	\end{equation}
   are to be said the anomalous cohomology spaces. 
    \end{definition}
The elements of $Z^r$ and $B^r$ will be called cocycles or closed elements and respectively coboundaries or exact elements in agreement with the language of mathematical literature.
    In order to identify the above spaces the several intermediate
    objects will be introduced and analyzed. The most important one is the relative complex.  

\subsection{Relative complex and bigrading}

In this subsection the relative complex (with respect to Cartan
subalgebra) will be analyzed in details with special attention paid to the structure of the corresponding differential.  Its relevance follows from well known observation that the cocycles outside the kernel of Cartan subalgebra  do not contribute to
cohomology\footnote{The Cartan subalgebra elements are (by assumptions made) diagonalizable. 
If $\Psi$ is a cocycle i.e. $D\,\Psi = 0$ and at the same time $L_i^\tot\,\Psi = \sigma_i\, \Psi\,$ for some $1\leq i \leq l\,$, then according to the remark made under (\ref{total_curvature}) one has $\,\Psi = D\,({\sigma_i}^{-1}\,b_i\,\Psi)$.}.
\\
The relative complex is, roughly speaking, defined as the kernel of Cartan subalgebra
elements $\{L_i^\tot,\}_ {i=1}^l\,$. Since the corresponding ghosts are of Cartan weights zero it is natural to get rid of them. To be more precise  
\begin{definition}[Relative complex]\ \\
\\ 
    The differential space
    $$
    \label{relative_complex}
    (\,\mathcal{A}_{\rm rel}\,,\,D\mid_{{\mathcal{A}}_{\rm rel}}\,)\;\;\;{\rm where}\;\;\;{\mathcal{A}}_{\rm rel} =
   \left\{\;\Psi \in \mathcal{A}\;;\;\; L_{H}^\tot\Psi = 0\;,\;\;H \in \mathfrak{h}\;,
   \;\;b_i\Psi = 0\;,\;\; i = 1 \ldots l\;\right\}\;,
    $$
   is to be said the anomalous relative complex.
\end{definition}
    The differential in the relative complex will be denoted by
    $D_{\rm rel}$.
    Since the ghosts of Cartan subalgebra are absent in
    $\mathcal{A}_{\rm rel}$ it is natural to change the grading of the
    relative cochains. The space of relative complex will be split
    into eigensubspaces corresponding to integral eigenvalues
    of relative ghost number operator ${\rm gh}_{\rel}$ introduced in
    (\ref{sl_2}). This amounts to the shift in the degree of
    cochains $r \rightarrow r + \frac{1}{2}l$, so that the
    elements which stem from ghost vacuum (\ref{vacuum}) are of
    relative degree zero. Consequently one has the decomposition:
    \begin{equation}
    \label{relative_complex_dec}
    \mathcal{A}_{\rm rel} = \bigoplus_{r=0}^m\,\mathcal{A}_{\rm rel}^r\;\;\;{\rm
    and}\;\;\;
    D_{\rm rel} : \mathcal{A}_{\rm rel}^r \rightarrow \mathcal{A}_{\rm rel}^{r+1}\;.
    \end{equation}
	It is worth to stress that the relative grading coincides with that defined by the 	
	weights with respect to $sl(2)$ Lie algebra of (\ref{sl_2_com}).\\
\\
Analogously to (\ref{anom_coho}) one introduces the relative cohomology spaces.
\begin{definition}[Relative cohomologies]\ \\
\\ 
    The quotient spaces
    \begin{equation}
    \label{relative_coho}
    H^r_{\rm rel}  = \frac{Z^r_{\rm rel}}{B^r_{\rm rel}}\;,\;\;\;\;\;\;Z^r_{\rm rel} = \ker D_{\rm rel}\! \mid_{\mathcal{A}^r_{\rm rel}}\;,\;\;B^r_{\rm rel}= {\rm im}\;
    D_{\rm rel}\! \mid_ {\mathcal{A}^{r-1}_{\rm rel}}
	\end{equation}
   are to be said the anomalous relative cohomology spaces. 
    \end{definition}
     It is not difficult to find  explicit expression for relative differential
     $D_{\rm rel}$ in terms of ghost modes and its relation with
     differential of the full complex.
    Using the
    definitions (\ref{koszul_diff_n}), (\ref{koszul_n}) and (\ref{total_l})
    together with (\ref{total_diff}) by simple separation of
    all the terms containing the ghosts corresponding to Cartan subalgebra elements,
     one may directly obtain:
    \begin{eqnarray}
    \label{relative_rel}
    D &=& D_{\rm rel} + \sum_{i=1}^l\,c^i\,L_i^\tot \,+ \,\sum_{i=1}^l\,M^i\,
    b_i\;,\;\;{\rm where}\;\; M^i = \{\,D\,,\,c^i\,\} =
    \sum_{\alpha > 0}\,c^{-\alpha}\,c^{\alpha}\,H_{\alpha}^i\;,\\
    \label{relative_modes}
    D_{\rm rel} &=& - \frac{1}{2}\sum_{\alpha\,\beta \in
    R}\,N_{\alpha\,\beta}\,c^{-\alpha}\,c^{-\beta}\,b_{\alpha+\beta}
    + \sum_{\alpha \in R}\,c^{-\alpha}\,\mathcal{L}_{\alpha}\;.
    \end{eqnarray}
The formulae (\ref{relative_rel}) establishes the rationship of  $D_{\rm rel}$ with absolute differential $D$. It appears to be 
 useful for reconstruction of absolute classes of (\ref{anom_coho}) out of relative ones.\\
\\
    The space $\mathcal{A}_{\rm rel}$ of relative complex (and the total space
$\mathcal{A}$ as well) admits richer grading structure than that
introduced by the eigenvalues of relative ghost number operator
${\rm gh}_{\rm rel}$. From the definition (\ref{sl_2}) it follows that the
operator ${\rm gh}_{\rm rel}$ is the difference of the  operators counting
the degrees of the cochains in  ghosts and anti-ghosts separately:
    \begin{equation}
    \label{ghost_dec}
    {\rm gh}_{\rm rel} = \overline{\textrm{gh}} - \textrm{gh}
    \;,\;\;\;{\rm where}\;\;\;\;
    \overline{{{\textrm{gh}}}} = \sum_{\alpha > 0}\, c^{-\alpha}\,b_{\alpha}
    \;\;\;{\rm and}\;\;\;
    \textrm{gh} = \sum_{\alpha > 0}\,b_{-\alpha}\,c^{\alpha}\,\;.
    \end{equation}
From the above expressions and  definitions of (\ref{sl_2}) it follows that one has
	\begin{equation}
	\label{partial_ghost_relations}
	[\,\overline{\textrm{gh}}\,,\,J^{\pm}\,] = \pm\, J^{\pm}\;,\;\;
	[\,{\textrm{gh}}\,,\,J^{\pm}\,] = \mp\,J^{\pm}\,, 
	\end{equation}
and consequently the total degree operator 
	\begin{equation}
	\label{degree_relations}
	{\rm deg} = \overline{\textrm{gh}} + \textrm{gh}\;\;\;\; {\rm is\; central}\;\;\;\;
	[\,{\textrm{deg}}\,,\,J^{\pm}\,] = 0\,, 
	\end{equation}
with respect to $sl(2)$ Lie algebra of (\ref{sl_2_com}). \\
The operators  (\ref{ghost_dec})  play  an important role in the identification of the harmonic elements of the appropriate Laplacians. At the moment they allow one to split the anomalous complex into bigraded subspaces as it is done in \cite{Wells}. \\
\\
Any element $\Psi^r \in \mathcal{A}_{\rm rel}^r$ can be decomposed into
bi-homogenous components $\Psi_q ^p$ such that
    $\overline{\textrm{gh}}\,\Psi_q ^p = p\,\Psi_q ^p$
    and
    $\textrm{gh}\,\Psi_q^p = q\,\Psi_q ^p\,$.
    Hence the decomposition of (\ref{relative_complex_dec}) can be made more
    subtle, namely:
    \begin{equation}
    \label{bidegree_dec}
    \mathcal{A}_{\rm rel}^r\, =
    \bigoplus_{p,q\;\;p-q = r}\,\mathcal{A}^p_q\;.
    \end{equation}
It appears that the bidegree decomposition of (\ref{bidegree_dec}) corresponds to the appropriate split of the relative differential $D_{\rm rel}$ - otherwise it would not be mentioned as irrelevant for the structure of the complex. 
In fact one is in a position to prove the following
    \begin{proposition}\
    \begin{enumerate}
\item
    The relative differential splits into bihomogeneous
    components:
    \begin{equation}
    \label{rozklad}
    D_{\rm rel} = \overline{\mathcal{D}} + \mathcal{D}\;,\;\; {\rm where
    }\;\;\;\;\;\;
    \overline{\mathcal{D}} : \mathcal{A}^p_q \rightarrow
    \mathcal{A}^{p+1}_q\;\;\;{\rm and}\;\;\;
    \mathcal{D} : \mathcal{A}^p_q \rightarrow
    \mathcal{A}^p_{q-1}\;,\;\;\;\;\;\;\;
    \end{equation}
\item
    The components are nilpotent and anticommute on the space of anomalous cochains:
    \begin{equation}
    \label{dede_bar}
    \overline{\mathcal{D}}^2 = 0 =
    \mathcal{D}^2\;,\;\;
    {and}\;\;\;\;\;\;\;\;\overline{\mathcal{D}}\mathcal{D} +
    \mathcal{D}\overline{\mathcal{D}} =\,-\, J^+\;\;
    (\;\equiv 0 \;{\rm on}\;\mathcal{A}\;)\;.\;\;\;\;
    \end{equation}
    \end{enumerate}
    \end{proposition}
{\it Proof:}\\
     {\it 1}. For the proof of the decomposition of
relative differential one should extract from
(\ref{relative_modes}) the components of bidegree $(1,0)$ and
$(0,-1)$. The first one denoted by $\overline{\mathcal{D}}$ raises
the eigenvalue of $\overline{\textrm{gh}}$ by $+1$ while the
second, $\mathcal{D}$
lowers the eigenvalue of $\textrm{gh}$ by $-1$. 
A straightforward calculation gives the following operator of
bi-degree $(1,0)$:
    \begin{eqnarray}
    \label{d_bar}
    \overline{\mathcal{D}} &=& \sum_{\alpha >0}\,c^{-\alpha}\,\mathcal{L}_{\alpha}
    + \overline{\partial} +
    \sum_{\alpha>0}\,c^{-\alpha} t_{\alpha}\;,\;\;
    {\rm where}\;\;\;\\
    \nonumber
    \overline{\partial} &=& - \frac{1}{2}\sum_{\alpha ,\,\beta>0}
    \,N_{\alpha\,\beta}c^{-\alpha}c^{-\beta}b_{\alpha + \beta}
    \;\;\;{\rm and}\;\;\;\;
    t_{\alpha} = - \sum_{\beta >
    \alpha}\,N_{\alpha\,-\beta}\,c^{\beta}b_{\alpha - \beta}\;.
    \end{eqnarray}
    Note that $\overline{\partial}$ is nothing else than the canonical
differential of nilpotent subalgebra $\mathfrak{n}_+ \subset
\mathfrak{g}$. The operators $t_{\alpha}$ describe the cross action
of $\mathfrak{n}_+$ on $\mathfrak{n}_-$.
    For the part of bi-degree $(0,-1)$ one similarly gets:
    \begin{eqnarray}
    \label{d_nie_bar}
    \mathcal{D} &=& \sum_{\alpha >0}\,c^{\alpha}\,\mathcal{L}_{-\alpha}
    + {\partial} +
    \sum_{\alpha>0}\,c^{\alpha} t_{-\alpha}\;,\;\;
    {\rm where}\;\;\;\\
    \nonumber
    {\partial} &=& - \frac{1}{2}\sum_{\alpha ,\,\beta>0}
    \,N_{\alpha\,\beta}c^{\alpha}c^{\beta}b_{-\alpha -\beta}
    \;\;\;{\rm and}\;\;\;\;
    t_{-\alpha} = - \sum_{\beta >
    \alpha}\,N_{\alpha\,-\beta}\,c^{-\beta}b_{\beta - \alpha}\;,
    \end{eqnarray}
    with ${\partial}$ being the differential of complementary $\mathfrak{n}_-$
    subalgebra and $t_{-\alpha}$ describing adjoint cross action
    of $\mathfrak{n}_-$ on $\mathfrak{n}_+$.\\
    {\it 2}. Using the equation (\ref{relative_rel}) one obtains $D_{\rm rel}^2 = D^2 - \sum_i M^i
    L_i$. The last term is zero on relative complex i.e. $D_{\rm rel}^2 =
    - J^+$. Taking into account the bidegree decomposition of $D_{\rm rel}$
    one obtains:
    $
    \overline{\mathcal{D}}^2 + \mathcal{D}^2+
    ({\mathcal{D}}\overline{\mathcal{D}}+
    \overline{\mathcal{D}}\mathcal{D}) = - J^+
    $.
    The first two terms are of bidegrees $(2,0)$ and $(0,-2)$
    respectively and they must vanish separately. The third one is
    of the type $(1,-1)$ hence it must be equal to the
    curvature.$\;\;\;\blacksquare$\\
\\
    It is important to remark that the differentials present in the decomposition of
    relative differential are related by
    conjugation defined in (\ref{conj_rules}): $(\mathcal{D})^\ast =
    \overline{\mathcal{D}}$.\\
    Although, according to (\ref{dede_bar}), the nilpotent differentials $\overline{\mathcal{D}}$
    and $\mathcal{D} $ anticommute on the anomalous complex $\mathcal{A}$ it
    will appear important to know the expression for
    $
    ({\mathcal{D}}\overline{\mathcal{D}}+
    \overline{\mathcal{D}}\mathcal{D})
    $
    outside this space.\\
    \\
    The bigraded structure of relative anomalous complex can be
    summarized in the form of the following diagram:
    \begin{equation}
    \label{diagram}
    \begin{CD}
    @.      \vdots  @.  \vdots  @.  \\
    @.      @VV{\cal D}V            @VV{\cal D}V        @.      \\
\cdots @>{\overline{{\cal D}}}>>    {\cal{A}}^p_{q}
@>{\overline{{\cal D}}}>>
{\cal{A}}^{p+1}_{q} @>{\overline{{\cal D}}}>>   \cdots\\
@.      @VV{\cal D}V            @VV{\cal D}V        @.      \\
\cdots @>{\overline{{\cal D}}}>>    {\cal{A}}^p_{q-1}
@>{\overline{{\cal
D}}}>>   {\cal{A}}^{p+1}_{q-1}   @>{\overline{{\cal D}}}>>   \cdots\\
@.      @VV{\cal D}V            @VV{\cal D}V        @.      \\
    @.      \vdots  @.  \vdots  @.
    \end{CD}
    \end{equation}
Since both $\overline{{\cal D}}$ and ${\cal D}$ are nilpotent one
may introduce the bigraded cohomology spaces:
    \begin{eqnarray}
    \label{bigraded_coho}
{\overline{\cal{H}}}^p_q  &=&
\frac{\overline{\cal{Z}}^p_q}{\overline{\cal{B}}^p_q}\;,\;\;\;\;\;\;\overline{\cal{Z}}^p_q
= \ker \overline{\cal{D}}\!
\mid_{\mathcal{A}^p_q}\;,\;\;\overline{\cal{B}}^r = {\rm im}\;
    \overline{\cal{D}}\! \mid_ {\mathcal{A}^{p-1}_q}
    \;\;\;\;{\rm and}\\
    \nonumber
    {\cal{H}}^p_q  &=& \frac{{\cal{Z}}^p_q}{{\cal{B}}^p_q}\;,\;\;\;\;\;\,{\cal{Z}}^p_q =
    \ker {\cal{D}}\!\mid_{{\mathcal{A}}^p_q}
    \;,\;\;
    {\cal{B}}^p_q = {\rm im}\;
    {\cal{D}}\!\mid_{{\mathcal{A}}^p_{q+1}}\;,
    \end{eqnarray}
which in contrast to the considerations on  K\"{a}hler geometry \cite{Wells} will not appear to be very useful in the identification of the cohomologies of the anomalous complex. The obstructions will be indicated below.

    \subsection{K\"{a}hler pairings}

The operations which do not preserve the kernel of the curvature operator i.e. do not act inside the space (\ref{relative_complex}) of relative anomalous complex are introduced in this chapter. Hence it is justified to go back to full differential space $\mathcal{C}(V)$ and especially its relative counterpart:
\begin{equation}
\label{relative_differential_space}
\mathcal{C}_{\rm rel}(V) :=  \left\{\;\Psi \in \mathcal{C}(V)\;;\;\; L_{H}^\tot\Psi = 0\;,\;\;H \in \mathfrak{h}\;,
   \;\;b_i\Psi = 0\;,\;\; i = 1 \ldots l\;\right\}\;.
\end{equation}
This space admits natural graded and bigraded structure which coincides with the ones defined in (\ref{relative_complex_dec}) and (\ref{bidegree_dec}) on anomalous complex:
$$
 \mathcal{C}_{\rm rel}(V) = \bigoplus_{r=- m}^m\,\mathcal{C}_{\rm rel}^r(V)\;,\;\;\;
 \mathcal{C}_{\rm rel}^r(V)\, =
    \bigoplus_{p,q\;\;p-q = r}\,\mathcal{C}^p_q(V)\;.
$$
The K\"{a}hler pairing on $\mathcal{C}_{\rm rel}(V)$ is appropriately induced by the corresponding Hodge - star conjugation.  The star operation is uniquely defined by the linear extension of the following rule:
    \begin{eqnarray}
    \label{hodge_star}
    &{{\star}}&\!\!\!\!\omega \otimes \varphi = \omega \otimes \varphi\;,\;\; \varphi \in V
\;\;\;{\rm and\;for \; the \; homogeneous\;
    elements\;}\;\;\;\\
    \nonumber
    &{{\star}}& \!\!\!\!(b_{-\alpha_1}\ldots
    b_{-\alpha_k}\,c^{-\beta_{1}}\ldots\,c^{-\beta_s}\omega \otimes \varphi)
    =(-1)^{(l+1)(k+s)+ ks}\prod_{i\,,j = 1}^{k\,,s}\,\frac{r_{\alpha_j}}{r_{\beta_i}}\;
   b_{-\beta_{1}}\ldots\,b_{-\beta_s}\, c^{-\alpha_1} \ldots
    c^{-\alpha_k}\;\omega \otimes \varphi\;,
    \end{eqnarray}
where $l = {\rm rank} (\mathfrak{g}) \,$. The sign factor depending on the rank of Lie algebra $\mathfrak{g}$ is introduced because of the presence of  Cartan subalgebra volume element in the normalization condition of the ghost vacuum (\ref{ghost_vacuum_norm}). It is necessary for positivity of the scalar product defined by $\star$. 
    It is clear that the mapping defined by the above rules is an
    isomorphism of the spaces of opposite bidegrees:
\begin{equation}
\label{star_map}
	\mathcal{C}^p_q(V) \stackrel{{{\star}}}{\rightarrow}
    \mathcal{C}^q_p(V)
	\;\;\;{\rm and\; extends\; to \;that \;of}\;\;\;
	\mathcal{C}^r_{\rm rel}(V)\stackrel{{{\star}}}{\rightarrow}
    \mathcal{C}^{\,-r}_{\rm rel}(V)\;.
\end{equation}   
 Note that in contrast to the standard definitions used in
    differential geometry, the star operation introduced above is
    idempotent ${{\star}}^2 = 1$. 
 The star operation induces a non degenerate and positive inner product on the whole
    space $\mathcal{C}_{\rm rel}(V)$ as well as in the anomalous complex $\mathcal{A}_{\rm rel}$ 
    \begin{equation}
    \label{inner_product}
    \langle\, \Psi\,,\,{\Psi}'\,\rangle =
    (\,{{\star}}\Psi\,,\upsilon (\mathfrak{h})
    \,{\Psi}'\,)\;,
    \end{equation}
    where $(\,\cdot,\cdot\,)$ denotes the original pairing on
    $\mathcal{C}(V)$ defined by (\ref{ghost_vacuum_norm}). \\
    The above scalar product defines the modified antiautomorphism of the operator algebra:
	\begin{equation}
	\label{dagger}
	\mathfrak{A} \;\rightarrow\; \mathfrak{A}^\dagger = 
	(-1)^{l\, {\rm deg}(\mathfrak{A})}\star \mathfrak{A} ^\ast \star\;,
	\end{equation}  
	where $ ^\ast $ denotes the original conjugation defined in (\ref{conj_rules}) and ${\rm deg}(\mathfrak{A})$ denotes the total degree in
$(b,c)$  generators and coincides with the weight of $\mathfrak{A}$ with respect to the central operator of (\ref{degree_relations}).\\
For the elementary ghost operators by straightforward calculation one obtains:
    \begin{equation}
    \label{kahler_conjugation}
    {c^{\alpha}}^\dagger \,=\,
    \frac{1}{r_{\alpha}}\,b_{- \alpha}\;,\;\;\;\;
    {b_{\alpha}}^\dagger \,=\,
    {r_{\alpha}}\,c^{- \alpha}\;,
    \end{equation}
    where $r_{\alpha}$ are the curvature coefficients introduced in
    (\ref{coefficients}).
    It should be stressed that star operation does not act inside
    of anomalous complex. It is clearly seen from the
    conjugation property of anomaly operator. As it follows from (\ref{sl_2}):
	 $({J^+})^\dagger = \star J^+ \star = J^-$, i.e. the $sl(2)$ -  highest weight vectors are transformed into the lowest weight ones under star action. Co nsequently the anomalous cochain may remain anomalous under star operation only if it is invariant with respect to $sl(2)$.\\
     The operator conjugated to the relative differential $D_{\rm rel}$ as well as those
	 conjugated to  (\ref{d_bar}) and (\ref{d_nie_bar}) are of special importance in the sequel. 
	Applying rules (\ref{conjugations_V}) and relations (\ref{kahler_conjugation})  one gets the following expression for the conjugate of the last one:
    \begin{eqnarray}
    \label{conjugate}
    \!\!\!\!\!\!\!\!\!\!{\mathcal{D}}^\dagger 
    =\frac{1}{2}\!\sum_{\alpha \,,\,\beta > 0}\!\!
    N_{\alpha \beta}
    \frac{r_{\alpha+\beta}}{r_\alpha \,r_{\beta}}c^{\alpha + \beta}b_{- \alpha}b_{-
    \beta}
    +\!\sum_{\alpha >0\,,\,\beta > \alpha}\!\!
    N_{-\alpha \beta}\frac{r_{\alpha-\beta}}{r_{\alpha}r_{\beta}}c^{\alpha-\beta}\,
    b_{- \alpha} b_{\beta}
     - \sum_{\alpha>0}\frac{1}{r_{\alpha}}\,b_{-
     \alpha}\,\mathcal{L}_{\alpha}\,.
    \end{eqnarray}
The corresponding explicit formulae for
    ${\overline{\mathcal{D}}}^\dagger$
    can be easily obtained by the original $^\ast$ conjugation of the
    above. From (\ref{star_map}) and (\ref{dagger}) it follows that 
	\begin{equation}
\label{daggered_action}
\overline{\mathcal{D}}^\dagger : \mathcal{C}^p_q(V) \rightarrow
    \mathcal{C}^{p-1}_q(V)\;\;\;{\rm and}\;\;\;
    \mathcal{D}^\dagger : \mathcal{C}^p_q(V) \rightarrow
    \mathcal{C}^p_{q+1}(V)\;.
	\end{equation}
Both conjugated operators are obviously nilpotent on the whole differential space and in addition
    ${{\mathcal{D}}}^\dagger{\overline{\mathcal{D}}}^\dagger+
    {\overline{\mathcal{D}}}^\dagger\mathcal{D}^\dagger = - J^-\,$. 
Their action does not preserve the anomalous complex i.e.  
	$\overline{\mathcal{D}}^\dagger\;(\,\mathcal{D}^\dagger\,) \;\mathcal{A}_{\rm rel}\, 
	\nsubseteq \,\mathcal{A}_{\rm rel}\, $ in general. This property of $\,^\dagger$ -  conjugated 
	differentials gets more detailed desctiption in the following
    \begin{lemma}
    \begin{equation}
    \label{j_plus_de}
    [\,J^+\,,\,{\mathcal{D}}^\dagger\,] = -
    \overline{\mathcal{D}}\;,
    \;\;\;\;
    [\,J^+\,,\,{\overline{{\mathcal{D}}}}^\dagger\,] =
    {\mathcal{D}}\;.
    \end{equation}
    \end{lemma}
    {\it Proof}: \\
    The proof of the first equality is obtained by direct
calculation with the help of cocycle property of $J^+$ explicitly
expressed in (\ref{cocycle}). The calculations are presented in
the {\it Appendix A}. The second equation is obtained by $^\ast$
conjugation of the former one.$\;\;\;\;\;\;\;\blacksquare$\\
\\
The above lemma says that although the property of cochains being
anomalous is not preserved by the action of $\,^\dagger$ -
conjugated differentials, the images of the space of  anomalous cocycles are contained in  the anomalous complex, more precisely: 
	\begin{equation}
\label{inclusions}
{\mathcal{D}}^\dagger \,\overline{\cal{Z}}^p_q \subset \mathcal{A}^p_{q+1}\;\;\;\;
{\rm and} \;\;\;\;{\overline{\mathcal{D}}}^\dagger\,{\cal{Z}}^p_q \subset
\mathcal{A}^{p-1}_{q}\;.
	\end{equation}
In order to obtain the similar characteristic of the action of ${D_{\rm rel}}^\dagger$  it is useful (as in \cite{Wells}) to introduce the
following operator associated to the relative differential:
    \begin{equation}
    \label{d_zero_ce}
    {D_{\rm rel}}^c = {\mathcal{D}}- \overline{{\mathcal{D}}}\;\;\;\;
    {\rm then\;obviously}\;\;\;\; D_{\rel}\,{D_{\rm rel}}^c + {D_{\rm rel}}^c D_{\rm rel} = 0\;,\;\;
    {D_{\rm rel}^c\,}^2 =  J^+\;.
    \end{equation}
From the above identites it follows that ${D_{\rm rel}}^c$  operator has (up to sign) the same curvature as $D_{\rm rel}$ and consequently it acts inside the space $\mathcal{A}_{\rm rel}$ of anomalous cochains.
From (\ref{j_plus_de}) one may immediately draw the following identities:
    \begin{equation}
    \label{jot_plus_de_zero_plus}
    [\,J^+\,,\,{D_0}^\dagger\,] = {D_0}^c \;\;\; {\rm and}
    \;\;\; [\,J^+\,,\,{{D_0}^c}^\dagger\,] = -  D_0\;.
    \end{equation}
Hence similarly to (\ref{inclusions}) one obtains:
\begin{equation}
\label{inclusions_1}
{D_{\rm rel}}^\dagger \; {Z_{\rm rel}^c}^r \subset \mathcal{A}_{\rm rel}^{r-1}\;\;\;\;
{\rm and}\;\;\;\;
{{D_{\rm rel}}^c}^\dagger \; {Z_{\rm rel}}^r \subset \mathcal{A}_{\rm rel}^{r-1}\;,
\end{equation}
where ${Z_{\rm rel}^c}^r = \ker {D_{\rm rel}}^c \mid_{\mathcal{A}_{\rel}^{r}}\,$.\\
 The above identities and relations appear to be crucial for finding the
    relationships between the Laplace operators corresponding to
    the differentials under consideration.
\subsection{Laplace operators}
    As in the standard nilpotent case \cite{Wells} one introduces the following family of Laplace operators determined by the pairing (\ref{inner_product}):
    \begin{eqnarray}
    \label{big_L}
    \bigtriangleup &=& D_{\rm rel}\,{D_{\rm rel}}^\dagger + {D_{\rm rel}}^\dagger D_{\rm rel}
    \;\;\;\;\;{\rm and}\;\;\;\;\;
    \bigtriangleup^c = {D_{\rm rel}}^c\,{{D_{\rm rel}}^c}^\dagger +
    {{D_{\rm rel}}^c}^\dagger {D_{\rm rel}}^c \;,\\
    \label{small_L}
    \Box &=& {\mathcal{D}}\,{\mathcal{D}}^\dagger  +
    {\mathcal{D}}^\dagger {\mathcal{D}}
    \;\;\;\;\;\;\;\;\;{\rm and}\;\;\;\;\;
    \overline{\Box}\;\; = \;\overline{{\mathcal{D}}}\,\overline{{\mathcal{D}}}^\dagger
    +
    \overline{{\mathcal{D}}}^\dagger\,\overline{{\mathcal{D}}}\;.
    \end{eqnarray}
All the operators introduced above are ghost number neutral and are self-adjoint with respect to the positive scalar product (\ref{inner_product}) on the relative differential space $\mathcal{C}_{\rm rel}(V)$ and relative anomalous complex 
$ \mathcal{A}_{\rm rel}\,$. Hence (under some analytical asuumptions on the $\mathfrak{g}$-module $V$ which will not be precised here) they are all diagonalizable.
From the definitions (\ref{big_L}) and (\ref{small_L}) it follows that their spectra are non-negative.  \\ However it is not yet clear that they do admit a common system of eigenvectors. The essential step towards this statement is presented below. The Laplace operators introduced above are not independent. In the  nilpotent case they are all proportional \cite{Wells}. In the anomalous case considered here some relations get modified.
    They are all presented in the following
    \begin{proposition}\
    \begin{enumerate}
    \item
    The Laplace operators $\bigtriangleup\;$, $\bigtriangleup^c$
    and $\;\Box\;$, $\overline{\Box}$ do satisfy the following
    relations:
    \begin{equation}
    \label{relations_shared}
    \bigtriangleup = \Box + \overline{\Box} = \bigtriangleup^c
    \;\;\;\;\;
    {\it and \;in\; addition}\;\;\;\;
    [\,J^\pm\,,\,\bigtriangleup\,] = 0\;.
    \end{equation}
    \item
    In contrast to $\bigtriangleup$ Laplace operator  the
    operators $\;\Box\;$ and $\;\overline{\Box}$ are not invariant with
    respect to $sl(2)$ algebra of (\ref{sl_2}). One has
    the following instead:
    \begin{equation}
    \label{relations_broken}
    [\,J^\pm\,,\,\Box\,] = \mp J^\pm\;\;,\;\;\;
    [\,J^\pm\,,\,\overline{\Box}\,] = \pm  J^\pm\;\;\;\; {\it and}
    \;\;\;\;
    [\,{\rm gh}_{\rm rel}\,,\,\Box\,] = 0 = [\,{\rm gh}_{\rm rel}\,,\,\overline{\Box}\,]\;.
    \end{equation}
    \item
    The $\Box\;$ and $\;\overline{\Box}$ Laplace operators are not
    equal, but are related by the following identity:
    \begin{equation}
    \label{broken_2}
     \Box = \overline{\Box} - {\rm gh}_{\rm rel}\;.
    \end{equation}
    \end{enumerate}
    \end{proposition}
    {\it Proof} :
    The proof is obtained by calculation similar to the
    one of \cite{Wells}.\\
    {\it 1}. Using the definitions of $\bigtriangleup$
    and $\Box,\overline{\Box}$
    together with decomposition (\ref{rozklad}) one obtains:
    $$
    \bigtriangleup = \Box + \overline{\Box} +
    {\mathcal{D}} \overline{\mathcal{D}}^\dagger +
    \overline{\mathcal{D}} {\mathcal{D}}^\dagger +
    {\mathcal{D}}^\dagger \overline{\mathcal{D}} +
    \overline{\mathcal{D}}^\dagger {\mathcal{D}}\;.
    $$
    It is then enough to check that the last four terms do vanish.
    According to (\ref{j_plus_de}) one obtains:
    $$
    \bigtriangleup - \Box - \overline{\Box} =
    (\;[\,J^+\,,\,\overline{{\mathcal{D}}}^\dagger\,]
    \overline{{\mathcal{D}}}^\dagger +
    \overline{{\mathcal{D}}}^\dagger
    [\,J^+\,,\,\overline{{\mathcal{D}}}^\dagger\,]\,)
    - (\,
    [\,J^+\,,\,{\mathcal{D}}^\dagger\,] {\mathcal{D}}^\dagger +
    {\mathcal{D}}^\dagger [\,J^+\,,\,{\mathcal{D}}^\dagger\,]
    \,)\;.
    $$
    Using nilpotency of both conjugated operators one immediately
    can see that both terms on the right hand side of the above
    equation do vanish independently and identically. The identity
    for $\bigtriangleup^c$ can be
    proved by exactly the same method. \\
    In order to check that the Laplace operator is invariant with
    respect to the $sl(2)$ algebra generated by anomaly it is
    enough to check that $\bigtriangleup$ commutes with $J^+$. The
    commutator with $J^-$ is obtained by $^\dagger$ conjugation.
    Using (\ref{d_zero_ce}) and (\ref{jot_plus_de_zero_plus} ) one
    immediately has:
    $$
    [\,J^+\,,\,\bigtriangleup\,] = D_{\rm rel}\,[\,J^+\,,\,D_{\rm rel}^\dagger\,]
    +
    [\,J^+\,,\,{D_{\rm rel}}^\dagger\,]\,D_{\rm rel} =
    (D_{\rm rel} {D_{\rm rel}}^c + {D_{\rm rel}}^c D_{\rm rel} ) = 0 \;.
    $$
    The fact that $\bigtriangleup$ is ghost number
    neutral follows from the above relations and the structural
    relations of (\ref{sl_2_com}) via Jacobi identity.\\
    {\it 2}. By direct calculation with the help of (\ref{j_plus_de}) and (\ref{dede_bar}) one gets:
    $
    [\,J^+\,,\,\Box\,] = [\,J^+\,,\,
    {\mathcal{D}}{\mathcal{D}}^\dagger + {\mathcal{D}}^\dagger
    {\mathcal{D}}\,] =
    {\mathcal{D}}[J^+\,,\,{\mathcal{D}}^\dagger\,] +
    [\,J^+\,,\,{\mathcal{D}}^\dagger\,] {\mathcal{D}} =
    - ({\mathcal{D}}\overline{{\mathcal{D}}}
    +\overline{{\mathcal{D}}}{\mathcal{D}}) =  J^+
    $. The
    commutator of $J^-$ with $\Box$ can be obtained by $^\dagger$
    conjugation while the commutation relations of $\overline{\Box}$
    with $J^\pm$ are obtained by $^\ast$ conjugation with the help of
    obvious properties $(J^\pm)^\ast = J^\pm$ and $\Box^\ast =
    \overline{\Box}$. According to the structural relations
    (\ref{sl_2_com}) of $sl(2)$ Lie algebra one writes
    $
    [\,\Box\,,\,{\rm gh}_{\rm rel}\,] = [\,\Box\,,[\,J^+\,,\,J^-\,]\,]
    $
    which is identically zero due to Jacobi identity and previous
    relations.\\
    {\it 3}. In order to prove the last identity it will be first shown
    that
    \begin{equation}
    \label{mixed_relations}
    {D_{\rm rel}}^\dagger\, {D_{\rm rel}}^c + {D_{\rm rel}}^c\,{D_{\rm rel}}^\dagger\,=\,- {\rm gh}_{\rm rel} = {{D_{\rm rel}}^c}^\dagger\, D_{\rm rel} +
    D_{\rm rel}\,{{D_{\rm rel}}^c}^\dagger\;.
    \end{equation}
    From (\ref{jot_plus_de_zero_plus}) one has
    $
    [\,J^-\,,\,{{D_{\rm rel}}^c}\,] = {D_{\rm rel}}^\dagger\,$ .
    Hence ${D_{\rm rel}}^\dagger\, {D_{\rm rel}}^c = [\,J^-\,,\,{{D_{\rm rel}}^c}\,]\,{D_{\rm rel}}^c
    =
    J^-\,{{D_{\rm rel}}^c\,}^2 - {D_{\rm rel}}^c J^-{D_{\rm rel}}^c
    $
    and analogously
    $
    {D_{\rm rel}}^c\,{D_{\rm rel}}^\dagger = {D_{\rm rel}}^c\,[\,J^-\,,\,{{D_{\rm rel}}^c}\,] =
    - {{D_{\rm rel}}^c}^2 J^- + { D_{\rm rel}}^c J^-{D_{\rm rel}}^c\,.
    $
    Adding up both expressions and taking into account (\ref{d_zero_ce}) one obtains:
    $
    {D_{\rm rel}}^\dagger\, {D_{\rm rel}}^c + {D_{\rm rel}}^c\,{D_{\rm rel}}^\dagger =
    [\,J^-\,,\,{{D_{\rm rel}}^c\,}^2\,] =  - [\,J^+\,,\,J^-\,]
    $,
    which due to (\ref{sl_2_com}) gives the first desired equality. The second one is obtained by
    $^\dagger$ conjugation of the first one.\\
    By (\ref{rozklad}) and (\ref{d_zero_ce}) one has
    $
    2\mathcal{D} = D_{\rm rel} + {D_{\rm rel}}^c\,.
    $
    Using definitions (\ref{big_L}) and relations
    (\ref{relations_shared}), (\ref{mixed_relations})
    one immediately obtains:
    $
    4\,\Box = 2\bigtriangleup +{ D_{\rm rel}}^\dagger\, {D_{\rm rel}}^c +
    {D_{\rm rel}}^c\,{D_{\rm rel}}^\dagger + {{D_{\rm rel}}^c}^\dagger\, D_{\rm rel} +
    D_{\rm rel}\,{{D_{\rm rel}}^c}^\dagger = 2(\Box + \overline{\Box}) - 2 \,{\rm
    gh}_{\rm rel}\;
    $, which gives (\ref{broken_2}).
    $\;\;\;\blacksquare$\\
    \\
    As it was already stated at the beginning of this section all
    the constructions related with K\"{a}hler pairings are consistent and
    all the related results above are valid on the whole relative differential
    space.\\
    From  proved relations (\ref{relations_shared}) and
    (\ref{relations_broken}) it follows that the actions of all introduced Laplace
    operators (\ref{big_L}) and (\ref{small_L}) preserve the anomalous
    relative complex. In addition they pairwisely commute i.e. they admit common 
	system of eigencochains. In the anomaly free case  the elements of their kernels fix the unique representatives of the anomalous cohomology classes. It will be proved that despite of the possible obstacles indicated above this statement remains true in the presence of non zero anomaly. For this reason  it is justified to remaind  the following
    \begin{definition}[Anomalous harmonic cochains]\ \\
\\
    The kernel of the Laplace operator
    \begin{equation}
    \label{harmonic}
    \mathfrak{H}(\bigtriangleup) := \ker
    \bigtriangleup\mid_{\cal{A}_{\rm rel}}
    \end{equation}
    is to be said the space of anomalous harmonic cochains.
    \end{definition}
	The space of harmonic cochains inherits the graded structure of $\cal{A}_{\rm rel}$ :  $\mathfrak{H}(\bigtriangleup) = \oplus_{r\geq 0} \mathfrak{H}^r(\bigtriangleup)$. From the definition (\ref{big_L}) and the positivity of the inner product on
	 $\mathcal{A}_{\rm rel}$ it immediately follows that
	$ \mathfrak{H}(\bigtriangleup) = \ker D_{\rm rel} \cap \ker {D_{\rm rel}}^\dagger\,$.\\
	One may analogously introduce the $\overline{\Box}$ and ${\Box}$-harmonic cochains.
	Since the operators $\overline{{\mathcal{D}}}$ and ${\mathcal{D}}$ are nilpotent on
	 the whole differential space it is by all means true that these harmonic elements 
	provide the representatives of the respective cohomology classes of the complexes
	 supported by total differential space.\\   
	The authors were however not in a position to prove the analogous statement (which
	 is probably not true) for the anomalous cohomologies of $\overline{{\mathcal{D}}}$
	 and ${\mathcal{D}}$ differentials. The obstacle that prevents one to establish
	 automatically the isomorphism between harmonic elements and cohomology classes has
	 its source in the cross relations of (\ref{inclusions}). \\
	Also in the case of relative anomalous complex the property of
	 $\bigtriangleup$ Laplace operator and $D_{\rm rel}$, that they act inside the
	 anomalous complex is not shared by $^\dagger$ - conjugated differentials and one has 
	the similar cross relations (\ref{inclusions_1}).  For this reason it is not 
	{\it a priori} clear that one 
	may establish the isomorphism between the spaces of anomalous harmonic elements
    and the anomalous cohomology spaces. Here the obstruction appears  to be 
	apparent however. \\
	In order to prove this last statement it is most convenient to use the spectral 
	decomposition of $\bigtriangleup$. The spaces of anomalous cochains decay into
	 mutually orthogonal subspaces of fixed eigenvalues
	\begin{equation}
	\label{spectral_dec}
	\mathcal{A}_{\rm rel}^r =  \mathfrak{H}^r(\bigtriangleup)\, \textstyle{\bigoplus}\, 
	\mathfrak{H}^r(\bigtriangleup)^\perp\;\;\; 
	 {\rm with}\;\;\;
	\mathfrak{H}^r(\bigtriangleup)^\perp = \textstyle{\bigoplus}_{\mu\, \in \,{{\rm spec}'(\bigtriangleup)}}\, \mathcal{A}_{\rm rel }^r(\mu)\;,
	\end{equation}
	where ${\rm spec(\bigtriangleup)}$ denotes the spectrum of $\bigtriangleup$ and 
	${{\rm spec}'(\bigtriangleup)} = {{\rm spec}(\bigtriangleup)} \setminus \{0\}$.\\
The essential step towards establishing the isomorphism of the space (\ref{harmonic}) of harmonic elements and relative cohomology  space $H_{\rm rel}$ is contained in the following 
	\begin{lemma}\ \\
\\
 Let
$$
 \Psi \in Z_{\rm rel}^r\;\;\; {\rm i.e.} \;\;\;D_{\rm rel} \Psi = 0\;\;\; {\rm and}
	 \;\;\;
	 \Psi = \Psi_0 + \Psi_+ \;\;\;{\rm where}\;\;\;\
	\Psi_0 \in \mathfrak{H}^r(\bigtriangleup)\;,\;\;\Psi _+\in \mathfrak{H}^r(\bigtriangleup)^\perp\,,
$$
then 
$$ 
\Psi_+ \in B_{\rm rel}^r\;\;\; {\rm i.e.} \;\;\;\Psi_+ = D_{\rm rel} \Phi\;\; 
	{\rm for\; some } \;\;\Phi \in \mathcal{A}_{\rm rel}^{r-1}\,.
$$
	\end{lemma}
 {\it Proof}:
Assume that $D_{\rm rel} \Psi = 0$ and $\Psi = \Psi_0 + \Psi_+$, where according to (\ref{spectral_dec}) $\Psi_+ = \sum_{\mu > 0} \Psi (\mu)$: 
	$\;\bigtriangleup \Psi (\mu) = \mu \Psi (\mu) $ with $D_{\rm rel} \Psi (\mu) = 0$. 
Then from (\ref{big_L}) one immediately obtains 
$\Psi (\mu) =  D_{\rm rel} \,(\,\mu^{-1} {D_{\rm rel}}^\dagger\, \Psi (\mu) ) $ and consequently 
$\Psi_+ = D_{\rm rel}\, (\,\sum_{\mu > 0}\mu^{-1}D_{\rm rel}^\dagger \Psi(\mu)\,)$. One has to prove that under assumptions made above the element $\Psi_+$ does belong to anomalous complex, which amounts to  
${D_{\rm rel}}^\dagger\, \Psi (\mu) \in \mathcal{A}_{\rm rel}^{r-1}(\mu) \, $ for all $\mu\, \in \,{{\rm spec}'(\bigtriangleup)}$ such that $\Psi(\mu)\neq 0$ .\\
From (\ref{jot_plus_de_zero_plus}) it follows that 
$J^+\,({D_{\rm rel}}^\dagger\, \Psi(\mu)) = [J^+\,,\,{D_{\rm rel}}^\dagger]\,\Psi(\mu) = {D_{\rm rel}}^c\, \Psi(\mu)\,$ and it will be shown that ${D_{\rm rel}}^c\, \Psi(\mu) = 0$, which guarantees that 
${D_{\rm rel}}^\dagger\, \Psi (\mu) \in \ker J^+$.\\
Due to (\ref{relations_shared}) one has $\bigtriangleup = \bigtriangleup^c$ and consequently also  
${D_{\rm rel}}^c\,{{D_{\rm rel}}^c}^\dagger + {{D_{\rm rel}}^c}^\dagger {D_{\rm rel}}^c \Psi(\mu) = \mu \Psi (\mu)\,$. Acting on both sides  with $D_{\rm rel}$ one obtains the following: 
$(D_{\rm rel} {D_{\rm rel}}^c\,{{D_{\rm rel}}^c}^\dagger + D_{\rm rel} {{D_{\rm rel}}^c}^\dagger {D_{\rm rel}}^c )\Psi (\mu) = 0\,$. 
Using $^\dagger$ - conjugate of the first relation in (\ref{jot_plus_de_zero_plus}):
${{D_{\rm rel}}^c}^\dagger = - [J^-\,,\,D_{\rm rel}] \,$, one may immediately write:
$$
0 = ( D_{\rm rel} {D_{\rm rel}}^c J^-D_{\rm rel} - D_{\rm rel} {D_{\rm rel}}^c D_{\rm rel} J^- + D_{\rm rel} J^- D_{\rm rel}{ D_{\rm rel}}^c - {D_{\rm rel}}^2 J^- 
{D_{\rm rel}}^c )\, 
\Psi(\mu)\,.
$$
The first  term does vanish as by assumption $D_{\rm rel} \Psi(\mu) = 0$. The third one is zero for the same reason if  the identity (\ref{d_zero_ce}): 
 $D_{\rm rel} {D_{\rm rel}}^c = - {D_{\rm rel}}^c D_{\rm rel}$ is taken into account. Using  this identity  once again and taking into account that, by assumption,  
$J^+ \Psi(\mu) = 0\,$ one may rewrite the above in the following form:
$$
0 = ( {D_{\rm rel}}^c {D_{\rm rel}}^2 J^- - {D_{\rm rel}}^2 J^- {D_{\rm rel}}^c) \Psi(\mu) = 
( {D_{\rm rel}}^c [J^+,J^-] - J^+ J^- {D_{\rm rel}}^c ) \Psi(\mu) \,.  
$$
Since $[ J^+,{D_{\rm rel}}^c ] = 0\,$ the product $J^+ J^-$ in the last term can be replaced by 
commutator $[J^+,J^-] = {\rm gh}_{\rm rel}$, which finally gives 
$$
0 = (\,[{D_{\rm rel}}^c,{\rm gh}_{\rm rel}]\,) \Psi(\mu) = - {D_{\rm rel}}^c\, \Psi(\mu)\,. 
$$
Hence ${D_{\rm rel}}^\dagger\, \Psi(\mu) \in \mathcal{A}_{\rm rel}^{r-1}(\mu)\,$. $\;\;\;\;\blacksquare$\\
\\
In fact, a bit more than stated in the above {\it Proposition} is proved. It was shown that ${D_{\rm rel}}^\dagger$ serves as contracting homotopy for anomalous cocycles outside the kernel of $\bigtriangleup\,$. This result is far from being obvious as  ${D_{\rm rel}}^\dagger$ does not preserve the kernel of $J^+$. \\
From the above proposition one may draw the following identification of cohomology spaces and an additional characterization of cocycles at relative ghost number zero, namely
\begin{corollary}\ 
\begin{enumerate}
\item
Every realtive cohomology class  has the unique harmonic representative:
\begin{equation}
\label{harm_rep_rel}
H_{{\rm rel}} ^r \simeq \mathfrak{H}^r(\bigtriangleup)\;.
\end{equation}
\item
Every cocycle of relative ghost number zero is harmonic:
\begin{equation}
\label{cocycle_harmonic}
Z^0_{{\rm rel}} = \mathfrak{H}^0(\bigtriangleup) = H_{\rm rel}^0\;.
\end{equation}
\end{enumerate}
\end{corollary}
{\it Proof}:\\
{\it 1.} Define the mapping $H_{\rm rel} ^r \ni [\Psi] \rightarrow h([\Psi]) 
	= \Psi_0 \in \mathfrak{H}^r(\bigtriangleup)\,$, where $\Psi_0$ denotes the the harmonic component of $\Psi\,$. The map is well defined as if $B_{\rm rel}^r \ni\Psi = D_{\rm rel} \Phi\,$, then 
for $\Psi_0 = (D_{\rm rel} \Phi)_0\,$ one obtains $\|\Psi_0\|^2 = \langle (D_{\rm rel} \Phi)_0\,,\,(D_{\rm rel}\Phi)_0\rangle = \langle D_{\rm rel} \Phi\,,\,(D_{\rm rel}\Phi)_0\rangle = \langle\Phi\,,\,D_{\rm rel}^\dagger\,(D_{\rm rel}\Phi)_0\rangle = 0\,$. The second equality was written because the decomposition of (\ref{spectral_dec}) is ortogonal, while the last one is true since harmonic elements are $D_{\rm rel}^\dagger$-closed. Hence $\Psi_0 = 0\,$. Every harmonic element is closed and consequently the  harmonic projection is onto. From the {\it Lemma} above it directly follows that it is also injective.
 \\
{\it 2.} Let $\Psi \in Z_{\rm rel}^0\,$:  $D_{\rm rel}\Psi = 0\,$. The cocycle can be decomposed according to (\ref{spectral_dec}): $\Psi = \Psi_0 + \Psi_+\;,\;\; \Psi_+ = \sum_{\mu >0}\Psi(\mu)\,$. It will be shown that  $\Psi_+ = 0\; $. In the the proof of the lemma above it was demonstrated that 
 first of all $\Psi_+ = D_{\rm rel}({D_{\rm rel}}^\dagger \Phi)$ for some $\Phi \in \mathcal{A}_{\rm rel}^0$ and moreover ${D_{\rm rel}}^\dagger \Phi \in \ker J^+\,$. Since ${D_{\rm rel}}^\dagger \Phi\,$ is of ghost degree $-1$ and anomalous at the same time, it must vanish identically. Hence $\Psi_+ = 0\,$.
$\;\;\;\;\blacksquare$ \\
It is important to stress that the space (\ref{harmonic}) of harmonic elements admits richer grading structure than that given by relative ghost number. Due to relations (\ref{relations_shared}) and (\ref{partial_ghost_relations} - \ref{degree_relations}) one may define the bi-degree decomposition
\begin{equation}
\label{bidegree_harm}
 \mathfrak{H}^r(\bigtriangleup) = \bigoplus_{p\geq 0} \mathfrak{H}^{r+p}_p(\bigtriangleup)\;,
\end{equation}
which means that every component of fixed bidegree of harmonic element is harmonic too. The subspace $\mathfrak{H}^{r}_0(\bigtriangleup)$ of lowest bidegree in the decomposition (\ref{bidegree_harm}) will be called the root component of $\mathfrak{H}^r(\bigtriangleup)$.\\
It is now straightforward to show the vanishing theorem for realtive anomalous cohomologies. 
\begin{proposition}[Vanishing theorem]\
\begin{enumerate}
\item The relative cohomology classes of positive ghost number are zero:
\begin{equation}
\label{vanishing}
H_{\rm rel}^r = 0 \;;\;\;\; r>0\,.
\end{equation}
\item
The root component of the cohomology at ghost number zero is isomorphic with the space of Borel subalgebra $\mathfrak{b}_+$-invariant elements of  $V$: 
\begin{equation}\
\label{zero_coh}
  \mathfrak{H}^{0}_0(\bigtriangleup) \simeq  V({\,\mathfrak{g},\chi })\;, 
\end{equation}
where  $V({\,\mathfrak{g},\chi })$ is that of (\ref{kernel}).
\end{enumerate}
\end{proposition}
{\it Proof:}\\
{\it 1.}From the statements above it follows that the harmonic cocycles do represent the relative cohomology classes in the unique way. Hence it is enough to show that $\mathfrak{H}^r(\bigtriangleup) = 0$ for $r>0$. From the relations (\ref{relations_shared}) and (\ref{broken_2}) it follows that $\bigtriangleup = 2\Box + {\rm gh}_{\rm rel}$. Then the equation $\bigtriangleup \Psi^r =0\;{\rm for}\;\Psi^r \in \mathcal{A}^r_{\rm rel} $ implies $\Box \Psi^r = - \frac{r}{2}\Psi^r$ which contradicts the evident property of $\Box$ being non negative if $r>0$.\\
{\it 2.} Let $\Psi_0^0$ be the harmonic element of bidegree $(0,0)$ i.e. of the form: $\Psi_0^0 = \omega\otimes \varphi$ where $\omega$ denotes the ghost vacuum (\ref{ghost_vac}) and $\varphi \in V$. Due to (\ref{cocycle_harmonic}) it must be a cocycle: $D_{\rm rel} \Psi_0^0 = 0$ which amounts (by (\ref{d_bar})) to $\mathcal{L}_{\alpha}\varphi = 0,\; \alpha > 0\,$. This property together with $\mathfrak{h}$ invariance of relative elements imply that $\varphi$ is $\mathfrak{b}_+$ invariant.
$\;\;\;\blacksquare$\\
The above statement does not exclude that there are non zero contributions of higher bidegrees to $H^0_{\rm rel} \sim \mathfrak{H}^{0}(\bigtriangleup)$. This problem is left open.\\
In fact one may suspect that stronger result than that of (\ref{zero_coh}) is true. Comparing the commutation relations  (\ref{relations_shared}) and (\ref{relations_broken}) of Laplace operators with $sl(2)$ generators with those of (\ref{partial_ghost_relations}) and (\ref{degree_relations}) one may guess that  $\overline{\Box}\,$, $\Box\,$ and $\bigtriangleup\,$ contain additively the corresponding $\overline{\rm gh}\,$, ${\rm gh}$ and ${\deg}$  operators. 
If one would be able to show that the respective differences $\overline{\Box} - \overline{\rm gh}\,$, $\Box - {\rm gh}\,$ and $\bigtriangleup - \deg\,$ are non-negative operators the vanishing theorem for relative cohomologies  at any non zero bidegree would follow automatically enhancing the identification given in  (\ref{zero_coh}).\\
Using the family $\{D_{\alpha}\}_{\alpha\in R}$ of operators (\ref{D_alpha_tot}) introduced in the {\it Appendix B} with crucial property ${D_{\alpha}}^\dagger = - D_{-\alpha}$
one is in a position to formulate  the the following  
\begin{proposition}[Conjecture]\ \\
\\
The Laplace operators are of the following form:
\begin{equation}
\label{lap_str}
	\Box = \mathfrak{K} + {\rm gh}\;,\;\;\;\overline{\Box} = \mathfrak{K} + 
\overline{{\rm gh}}\;,\;\;\;\bigtriangleup = 2 \,\mathfrak{K} + \deg\;,
\end{equation}
where
\begin{equation}
\label{K_str}
	\mathfrak{K} =  - \sum_{\alpha >0}\,\frac{1}{r_\alpha}\,D_{-\alpha}\,D_{\alpha}\;.
\end{equation}
\end{proposition}
{\it Proof}:\\
It is enough to prove first equality above as the remaining ones follow directly from the formulae (\ref{relations_shared}) and (\ref{relations_broken}). The detailed calculations showing this very first equality might be true are presented in the {\it Appendix B}.
$\;\;\;\;\blacksquare ?$\\
It is worth to mention that on the elements which stem from the ghost vacuum i.e. those of total degree zero:  $\Psi = \omega \otimes \varphi\,$, the  Laplace operator $\bigtriangleup$ reduces to:
\begin{equation}
\label{Lap_red}
\bigtriangleup \simeq -2\,\sum_{\alpha>0}\, 
\frac{1}{r_\alpha} \,\mathcal{L}_{-\alpha}\,\mathcal{L}_{\alpha}\;.
\end{equation}
This expression can also be obtained by straightforward calculation independently of whether the conjecture is true or not.\\
\\
From the formulae (\ref{K_str}) it immediately follows that the operator $\mathfrak{K}$ is non-negative. Therefore from the above {\it Proposition} one may directly deduce the statement on vanishing of higher ghost number relative cohomology classes. Moreover it can be immediately shown that any non trivial cohomology class is represented by ghost free element of $\mathcal{A}$ subject to the conditions of (\ref{kernel}).\\
It is worth to mention that the Laplace operators corresponding to (\ref{lap_str}) were computed in the context of anomalous relativistic models of high spin particles (\cite{my}). There they appeared to serve as the operators defining the Lagrange densities implying (via Euler - Lagrange principle) the irreducibility equations (of Dirac type) for the relativistic fields carrying arbitrarily high spin.\\
The next section is devoted to identification of absolute cohomologies. It has to be stressed that the conjectured results will never be used.
\subsection{Absolute cohomologies}
 The vanishing theorem (\ref{vanishing}) allows one to determine easily the absolute cohomologies (\ref{anom_coho}) of the anomalous complex out of those of relative one. The result can be almost immediately read off from the general theory of spectral sequences \cite{BT}. In order to make the paper self-contained the way of reasoning for the present special case will be presented here in less abstract form. For this reason it is convenient to introduce some definitions and to fix some notation. \\
The spaces (\ref{asymm}) of absolute, $\mathfrak{h}$-invariant, anomalous 
cochains\footnote{As it was already mentioned the restriction of the complex to $\mathfrak{h}$-invariant subspace does not change the cohomology.}
can be reconstructed out of those of relative complex. In order to keep the consistency of the notation it is necessary to come back to the original grading of the absolute complex. According to the relation (\ref{sl_2}) between absolute ghost number operator and that of relative ghost number the subspace $\mathcal{A}^r\,, \;r\geq -\frac{1}{2}l\,$ of fixed (absolute) ghost number can be decomposed according to the content of Cartan subalgebra ghosts $\{\,c^i\,\}_{i=1}^l\,$:
\begin{equation}
\label{absolute_dec_0}
\mathcal{A}^r = 
\bigoplus_{{s=0}}^{m(l,r)}\,\textstyle{{\bigwedge}}^s\,\mathfrak{h}^\ast\,
\mathcal{A}_0^{R-s}\;,\;\;R=r+\textstyle{\frac{1}{2}}l\;,
\end{equation}
where $m(l,R)= \min \{l,R\}\,$ and $\bigwedge^s\,\mathfrak{h}^\ast$ is generated by all "clusters" of Cartan subalgebra ghosts of ghost number (degree) $s\,$.\\
Due to (\ref{absolute_dec_0}) any element of the space $\mathcal{A}^r$ can be expanded as a form from $\bigwedge\,\mathfrak{h}^\ast$ with coefficients from relative complex of appropriately adjusted ghost numbers:
\begin{equation}
\label{expansion}
\Psi^{-\frac{1}{2}l+R} = \sum_{s=0}^{m(l,R)}\,\sum_{\{\,I_s\,\}}\,
c^{I_s}\,\Psi_{I_s}^{R-s}\;,\;\;\; \Psi_{I_s}^{R-s} \in \mathcal{A}_{\rm rel}^{R-s}\;,
\end{equation}
where $\{\,I_s\,\}\,$ denotes the set of all monotonically ordered multiindices: 
$I_s = (i_1\ldots i_s)\,$ of length $s\,$ and $c^{I_s}= c^{i_1}\ldots c^{i_s}\,$ generate the basis elements of $\bigwedge^s\,\mathfrak{h}^\ast$.\\
It is clear that not all  the admissible relative degrees of the coefficients  (\ref{expansion}) are necessarily present (are non zero) in (\ref{expansion}). For this reason is convenient to introduce the the notion of the basic component of the absolute cochain. It is defined as non zero ingredient containing all summands of lowest relative ghost number in (\ref{expansion}).\\ 
More precisely: the element (\ref{expansion}) of $\mathcal{A}^r$ is said to {\it stem} from relative ghost number $r_0$ if and only if $\,\Psi^{R - s_0}_{I_{s_0}} \neq 0$ for some $I_{s_0}$ with $0\leq R-s_0 = r_0$ and $\Psi^{R - s}_{I_{s}} = 0$ for $s>s_0\,$.\\ 
Then the sum containing highest degree in Cartan subalgebra ghosts: 
 \begin{equation}
\label{root_abs}
\Psi^{-\frac{1}{2}l+R}_{\rm root} = \,\sum_{\{\,I_{s_0}\,\}}\,
c^{I_{s_0}}\,\Psi_{I_{s_0}}^{r_0}\;,\;\
\end{equation}
is called the  {\it basic component} of $\Psi^{-\frac{1}{2}l+R}\,$.
It is now possible to prove the following 
\begin{lemma}\ \\
\\
Any absolute cocycle which stems from positive relative ghost number is a coboundary. 
\end{lemma}
{\it Proof}:
According to (\ref{expansion}) any element $\Psi^{-\frac{1}{2}l+R}\,$ which stems from relative degree $r_0$ can be expanded as:
\begin{equation}
\label{expansion_1}
\Psi^{-\frac{1}{2}l+R}\,= \,\sum_{s=0}^{s_0}\,\sum_{\{I_{s_0-s}\}}\,
c^{I_{s_0-s}}\,\Psi^{r_0 +s}_{I_{s_0-s}}\;,\;\;\;s_0+r_0 = R\,.
\end{equation}
Using the formulae (\ref{relative_rel}) which gives the relationship of absolute differential $D$ with relative differential $D_{\rm rel}$ one can write the following equation for (\ref{expansion_1}) to be a cocycle:
\begin{equation}
\label{de=0}
0= D\Psi^{-\frac{1}{2}l+R}\,= \,\sum_{s=0}^{s_0} \,(\,\,(-1)^{(s_0-s)}\sum_{\{I_{s_0-s}\}}\,
c^{I_{s_0-s}}\,D_{\rm rel}\,\Psi^{r_0 +s}_{I_{s_0-s}}\, + \,
\sum_{\{I_{s_0-s}\}}\,\sum_{i=1}^l\,
[\,b_i,c^{I_{s_0-s}}\,\}\,M^i\Psi^{r_0 +s}_{I_{s_0-s}}\,\,)\;,
\end{equation}
where $[\,b_i,c^{I_{s_0-s}}\,\}$ denotes the (anti-) commutator for (odd) even $s_0-s$ and 
$M^i = \{\,D,c^i\,\}$ are explicitly given in (\ref{relative_rel}). It is worth to stress again that since the cochains are (by assumption) in the kernel of Cartan subalgebra elements the part 
$\sum_{i=1}^{l}\,c^i\,L_i^\tot$ of absolute differential was neglected.\\
The equation (\ref{de=0}) decays into the system of independent relations. The most important is the one for the basic component of $D\Psi^{-\frac{1}{2}l+R}\,$:
\begin{equation}
\label{de=0_root}
0 = \sum_{\{I_{s_0}\}}\,
c^{I_{s_0}}\,D_{\rm rel}\,\Psi^{r_0}_{I_{s_0}}\;\;\;{\rm implying}\;\;\;
 D_{\rm rel}\,\Psi^{r_0}_{I_{s_0}} = 0\;\;\;{\rm for\;all}\;\;\;I_{s_0}\;. 
\end{equation}
Due to vanishing theorem (\ref{vanishing}) for relative cohomologies there exsist the cochains $\{\,\Phi^{r_0-1}_{I_{s_0}}\,\}\,$ such that 
$\Psi^{r_0}_{I_{s_0}} = D_{\rm rel}\,\Phi^{r_0-1}_{I_{s_0}}\,$ for all $I_{s_0}\,$.
The "gauge" shift 
\begin{equation}
\label{shift}
\Psi^{-\frac{1}{2}l+R} \;\rightarrow \;\Psi^{-\frac{1}{2}l+R}\, - \,
D\,(\,( -1)^{s_0}\sum_{\{I_{s_0}\}}\,
c^{I_{s_0}}\,\Phi^{r_0-1}_{I_{s_0}}\,)\;, 
\end{equation}
kills the basic component  and one gets an equivalent cocycle which stems from at least relative ghost number $r_0+1\,$. It is now clear that proceeding by induction it is possible to obtain an equivalent cocycle of degree zero in Cartan ghosts: 
$\,\Psi^{-\frac{1}{2}l+R} \,\sim \,\Psi^{r_0+s_0}\,$. The equation $D\,\Psi^{-\frac{1}{2}l+R} = 0 \,$ amounts to $D_{\rm rel}\,\Psi^{r_0+s_0}= 0\,$ and is solved by 
$\Psi^{-\frac{1}{2}l+R} = D_{\rm rel}\,\Phi^{r_0+s_0 -1} = D\,\Phi^{r_0+s_0 -1}$ as $D$ and $D_{\rm rel}\,$ coincide on relative cochains.$\;\;\;\;\blacksquare$\\
\\
The above {\it Lemma} implies the following important
\begin{corollary}\ \\
\\
The absolute cohomology spaces of ghost number $r > \frac{1}{2}l\,$ are zero:
\begin{equation}
\label{abs_vanish}
H^r \,=\, 0 \;,\;\;\;r > \textstyle{\frac{1}{2}}l\;.
\end{equation}
\end{corollary}
{\it Proof:} 
As it can be easily seen from the expansion (\ref{expansion}) all the elements of $\mathcal{A}^r\,$ with $r > \frac{1}{2}l\,$ stem from positive relative ghost numbers and the conclusion of  {\it Lemma} applies automatically.
$\;\;\;\;\blacksquare$\\
\\
The situation is different for the ghost numbers within the range 
$-\frac{1}{2}l \leq r \leq \frac{1}{2}l\,$, but also here the result of {\it Lemma} appears to be helpful. 
\begin{proposition}[Absolute cohomology]\ \\
\\
The absolute cohomology spaces $H^r\,$ are non zero for $-\frac{1}{2}l \leq r \leq \frac{1}{2}l\,$ and
\begin{equation}
\label{absolute_classes}
H^{\,-\frac{1}{2}l+s} \,\simeq \,
\textstyle{\bigwedge}^s \,\mathfrak{h}^\ast\,\mathfrak{H}(\bigtriangleup)\;,\;\;\;0\leq s \leq l\;,
\end{equation}
where $\mathfrak{H}(\bigtriangleup)\,$ is the space of harmonic cochains.
\end{proposition}
{\it Proof}: Let $\Psi^r \in \mathcal{A}^{\,-\frac{1}{2}l+s}$ for some fixed $0\leq s \leq l\,$. According to (\ref{expansion_1}) this element can be expanded as:
\begin{equation}
\label{expansion_2}
\Psi^{r}\,= \,\sum_{s'=0}^{s}\,\sum_{\{I_{s-s'}\}}\,
c^{I_{s-s'}}\,\Psi^{s'}_{I_{s-s'}}\;.
\end{equation}
If the above cochain is a cocycle, then by {\it Lemma} it is cohomologous to the one which stems from relative ghost number zero. The expansion (\ref{expansion_2}) canbe limited to $s'=0$.\\ It will be shown that any absolute cocycle is equivalent to its basic component of ghost number zero which in turn, contributes to absolute cohomology non triviallly. The equation (\ref{de=0}) reduces to the following one:
\begin{equation}
\label{de_0_red}
0 = D \Psi^r = \,\,(-1)^{s}\sum_{\{I_{s}\}}\,
c^{I_{s}}\,D_{\rm rel}\,\Psi^{ 0}_{I_{s}}\, + \,
\sum_{\{I_{s}\}}\,\sum_{i=1}^l\,
\,M^i b_i\,c^{I_{s}}\,\Psi^{0}_{I_{s}}\;,\;\; r = -\frac{l}{2}l + s
\end{equation}
Simple reasoning shows that any absolute cocycle $\Psi^r$ is cohomologically equivalent to  $\tilde{\Psi} ^r$  - the one such that $\sum_{i=1}^l\,
\,M^i b_i\,\tilde {\Psi}^r = 0\,$. Indeed, making the "gauge shift": 
$$
\Psi^r \;\rightarrow \; \tilde{\Psi}^r = 
\Psi^r - (-1)^s D\sum_{i=1}^{l}\,c^i b_i \Psi^r
$$
one obtains a cocycle such that
\footnote{One should exploit the possibility that 
$\tilde{\Psi} ^r$ can be chosen to be $\mathfrak{h}$ - invariant.} $D\tilde{\Psi}^r = D_{\rm rel}\tilde{\Psi}^r$.\\
The equation (\ref{de_0_red}) implies $D_{\rm rel}\tilde{\Psi}^{0}_{I_{s}} = 0$.
for any component and the result follows. $\;\;\;\; \blacksquare$\\
It was proved that the absolute cohomology space (\ref{absolute_classes}) contains $2^{l}, \; l = {\rm rank \mathfrak{g}}\,$ copies of the relative cohomology spaces. This degeneracy is  necessary and in fact sufficient to induce a non degenerate pairing of cohomology classes from the one (\ref{ghost_vacuum_norm}-\ref{conj_rules}) introduced on total differential space at the very begining of the paper. On the other hand it would be significant to give the possible interpretation of this degeneracy within the physical context. The same remark concerns the possible degeneracy of the relative cohomology space  but here the situation is  more complicated as it is difficult to decide if the content of $H_{\rm rel}^0$ can be described by the copies of Gupta-Bleuler space of states (\ref{kernel}). It seems that these questions cannot be answered within the scope of general procedure presented in this article. For this reason   some  model with direct physical interpretation should be pushed through the formalism of anomalous cohomologies. This issue runs outside the scope of the paper.

\section{Concluding remarks}
 There is another complex associated with the anomalous system of constraints. The differential space
    \begin{equation}
    \label{polarized_complex}
    (\,\mathcal{P}\,,\,D\mid_{\mathcal{P}}\,)\;\;\;{\rm where}\;\;\;\mathcal{P} =
    \ker \,[\,\mathfrak{b}_+^\tot\,,\,D\,]\;,
    \end{equation}
    is to be said the polarized complex.
For the sake of simplicity of the notation, the mark of restriction
    $\,.\mid$ of the domain will be suppressed in the sequel.\\
From (\ref{total_curvature}) it follows that the space $\mathcal{P}$ of polarized complex consists of the cochains, which are anihilated by all $c^\alpha\,;\;\alpha > 0\,$ ghost operators. Hence it may be identified with 
$\bigwedge \mathfrak{b}_+^* \omega \otimes V\, $.\\
It is obvious that $(\,\mathcal{A}\,, D\,)$ is a complex. In the
case of $(\,\mathcal{P}\,, D\,)$ one has to pay a more of attention
in order to check that the definition (\ref{polarized_complex}) is
consistent. One may prove the following
    \begin{lemma}\ \\

    $(\,\mathcal{P}\,,D\,)\;$ is a subcomplex of
    $\;(\,\mathcal{A}\,,D\,)\;$ i.e.
    $\;\;
    \mathcal{P}\, \subset\, \mathcal{A}\;\;\;$ and $\;\;\;
    D\,\mathcal{P}\,\subset\,\mathcal{P}\;.
    $
    \end{lemma}
{\it Proof :} The property that
    $\mathcal{P}\, \subset\,\mathcal{A}\,$
    follows from the right hand side formulae of {\it Proposition
    1.2}. One simply has
    $D^2 = \sum_{\alpha > 0}\,
    c^{-\alpha}\,[\,L_{\alpha}^\tot\,,\,D\,]\,$ and the first inclusion is immediate.\\
    In order to prove the second one one should note that for any $x \in
    \mathfrak{b}_+\,$ the following identity is true
    $[\,L_x^\tot\,,D\,]\,D = [\,L_x^\tot\,,\,D^2\, ]$ on $\mathcal{P}$.
    Using the formulae for $D^2$ and Leibnitz rules one may write
    $
    [\,L_{\beta}^\tot\,,D^2\,] = \sum_{\alpha
    >0}\,[\,L_{\beta}^\tot\,,\,c^{-\alpha}\,]\,[\,L_{\alpha}^\tot\,,\,D\,]
    + \sum_{\alpha >0}\,c^{-\alpha}(\,[\,[\,L_{\beta}^\tot\,,L_{\alpha}^\tot\,]\,,\,D\,] +
    [\,L_{\alpha}^\tot\,,\,[\,L_{\beta}^\tot\,,\,D\,]\,]\,)\,.
    $
    The first  term does vanish on $\mathcal{P}$ by definition of
    polarized complex. The second one is zero for the same reason
    if one would take  into account the fact that $\mathfrak{b}_+$ is Lie
    subalgebra of $\mathfrak{g}\,$. Finally, due to (\ref{total_curvature})
    one is left with the identity
    $
    [\,L_\beta^\tot \,,D\,]\,D = \sum_{\alpha > 0}\,c^{-\alpha}
    \,<\chi +
    2\varrho\,,\,H_{\beta}>\,[\,L_{\alpha}^\tot\,,\,c^{\beta}\,]\,
    $
    on $\mathcal{P}$ for any $\beta > 0 $. The proof is finished
    by observation that $[\,L_{\alpha}^\tot\,,\, c^{\beta}\,]$ is
    proportional to $c^\gamma$ for some $\gamma >0$ which is zero
    on $\mathcal{P}$ - again due to right hand side identity of
    (\ref{total_curvature}).
    Hence $\mathcal{P}$ is stable under action of differential. $\;\;\blacksquare$\\
The polarized complex was not investigated in this paper for at least two reasons. First of all its definition is much less natural than the one of anomalous complex if not artificial. Secondly it is rather hopeless to expect that the cohomologies of polarized complex are vanishing as they do in anomalous case giving quite acceptable description of the Gupta-Bleuler "physical space" (\ref{kernel}).\\\\
It is not difficult to generalize the constructions of anomalous complexes relaxing the assumption that  $\chi$ is regular. In this case the anomaly gets singular and  there is non-abelian Lie algebra of first class constraints replacing  Cartan subalgebra. Its structure depends on the location of $\chi$ on the walls of Weyl chamber. Since the considerations become embroiled with Lie algebraic details in singular case the athors postponed a presentation of respective constructions and results to the future paper.\\
In many aspects the problem considered in this paper resembles the gauge theory with gauge symmetry broken by Higgs potential. It is then tempting to speculate on the possible physical interpretation of the degeneracy of the cohomology spaces in this context. Here the degeneracy may correspond to additional quantum numbers which automatically appear within the cohomological framework as strictly related to the original gauge symmetry. However this kind of speculations  becomes to close to science fiction at the present stage of considerations. 

 \eject
    \section*{Appendix A}
    \renewcommand{\theequation}{A.\arabic{equation}}
    \setcounter{equation}{0}
In order to prove the formulae (\ref{j_plus_de})  the commutation rule 
\begin{equation}
\label{jot_plus_be}
[J^+,b_\alpha] = - {\rm sign}(\alpha)\, r_{\alpha}\, c^\alpha\,,
\end{equation}
which immediately follows from the definition (\ref{sl_2}) and structural relations (\ref{clifford_r}) appears to be very helpful.
After obvious change of summation range (in the middle term of(\ref{conjugate})), the operator ${\mathcal{D}}^\dagger$  can be rewritten in the following form:
\begin{equation}
\label{conjugate_a}
{\mathcal{D}}^\dagger = 
\frac{1}{2}\!\sum_{\alpha \,,\, \beta > 0}\!\!
    N_{\alpha \beta}
    \frac{r_{\alpha+\beta}}{r_\alpha \,r_{\beta}}
	c^{\alpha + \beta}b_{- \alpha}b_{-\beta}
    +\!\sum_{\alpha \,,\,\beta > 0}\!\!
    N_{-\alpha\, \alpha+\beta}\frac{r_{\beta}}{r_{\alpha}r_{\alpha+\beta}}
    c^{-\beta}\,b_{- \alpha} b_{\alpha + \beta}
     - \sum_{\alpha>0}\frac{1}{r_{\alpha}}b_{- \alpha}\mathcal{L}_{\alpha} 
	\,.
\end{equation}
Denote the three terms obove as D.1, D.2 and D.3 respectively. For the simplest commutator one obtains:
\begin{equation}
\label{d.1}
[J^+, ({\rm D.3})] = - \sum_{\alpha >0}\frac{1}{r_{\alpha}}\,
[J^+, b_{- \alpha}]\,\mathcal{L}_{\alpha}
 = - \sum_{\alpha >0}c^{-\alpha}\,\mathcal{L}_{\alpha}\,. 
\end{equation}
 Due to (\ref{jot_plus_be}) the commutator of $J^+$ with middle part of (\ref{conjugate_a}) is given by
\begin{equation}
\label{d.2}
[J^+ ,{\rm D.2}] = \sum_{\alpha \,,\,\beta > 0}\!\!
    N_{-\alpha\, \alpha+\beta}\left(\frac{r_{\beta}}{r_{\alpha+\beta}} c^{-\beta}
c^{-\alpha}b_{\alpha + \beta} - 
\frac{r_{\beta}}{r_{\alpha}}c^{-\beta} b_{-\alpha} c^{\alpha + \beta} \right)\,.
\end{equation} 
Denote the first and the second term above by D.2.1 and D.2.2 respectively. 
The first one can be antisymmetrized to give:
\begin{equation}
\label{a.s}
{\rm D.2.1} =  \frac{1}{2} \sum_{\alpha \,,\,\beta > 0}\! 
\left(\, N_{ \alpha+\beta\;-\alpha}\frac{r_{\beta}}{r_{\alpha+\beta}}  -
N_{ \alpha+\beta\;-\beta}\frac{r_{\alpha}}{r_{\alpha+\beta}}\,
\right) 
c^{-\alpha} c^{-\beta}b_{\alpha + \beta} \;.
\end{equation}
Using the cocycle identity (\ref{cocycle}) one obtains finally:
\begin{equation}
\label{a.s.end}
{\rm D.2.1} = \frac{1}{2} \sum_{\alpha \,,\,\beta > 0}\!N_{ \alpha \beta}
c^{-\alpha} c^{-\beta}b_{\alpha + \beta} = - \;\overline{\partial}\;, 
\end{equation}
where $\overline{\partial}$ is that of (\ref{d_bar}) .\\
The commutator of $J^+$ with D.1 part of (\ref{conjugate_a}) can be calculated to be: 
\begin{equation}
\label{d.1.1}
[J^+,{\rm D.1}] = \frac{1}{2} \sum_{\alpha ,\,\beta > 0}\!N_{ \alpha \beta}
c^{\alpha + \beta}\left(\frac{r_{\alpha + \beta}}{r_{\beta}} c^{-\alpha}b_{-\beta} +
\frac{r_{\alpha+\beta}}{r_{\alpha}}b_{-\alpha}c^{-\beta}
\right) = \sum_{\alpha ,\,\beta > 0}\!N_{ \alpha \beta}
\frac{r_{\alpha + \beta}}{r_{\beta}}c^{\alpha + \beta}c^{-\alpha}b_{-\beta}\,.
\end{equation}
The last equality is written due to antisymmetry of $N_{\alpha \beta}$ structure constants. Adding  D.2.2 and the above expression one obtains:
\begin{equation}
\label{add_up}
[J^+,{\rm D.1}] + {\rm D.2.2} = 
\sum_{\alpha ,\,\beta > 0} 
\left(
N_{\alpha \beta}\frac{r_{\alpha + \beta}}{r_{\beta}} +
N_{ \alpha + \beta\;-\beta}\frac{r_{\alpha }}{r_{\beta}}
\right) 
c^{\alpha + \beta}c^{-\alpha}b_{-\beta}\,.
\end{equation}
Using again the cocycle property (\ref{cocycle}) one may write
\begin{equation}
\label{last}
{\rm A.8} =  \sum_{\alpha ,\,\beta > 0} 
N_{\alpha + \beta\;-\alpha}c^{\alpha + \beta}c^{-\alpha}b_{-\beta} = 
 - \sum_{\alpha >0} c^{-\alpha}
\left(
\sum_{\beta > 0} N_{\alpha + \beta\;-\alpha}c^{\alpha + \beta}b_{-\beta}
\right)\,, 
\end{equation}
which after natural change of summation range and with the use of the property 
$N_{-\alpha\,-\beta} =
    N_{\alpha\,\beta} $
of the structure constants in Chevalley basis gives:
$$
{\rm A.8} = - \sum_{\alpha >0} c^{-\alpha} t_{\alpha}\,,
$$
where $t_{\alpha}$ are that of (\ref{d_bar}). \\
Adding up all the calculated terms one obtains the expression for 
$ - \overline{\mathcal{D}}$ . 

 \section*{Appendix B}
    \renewcommand{\theequation}{B.\arabic{equation}}
    \setcounter{equation}{0}
In order to recover the form of Laplace operators one should start with some technical preparation. First of all, it is convenient to split the differentials $\mathcal{D}$ of (\ref{d_nie_bar}) and 
$\overline{\mathcal{D}}$ of (\ref{d_bar}) into purely ghost part and the one acting acting non-trivially on the tensor factor $V$:
\begin{eqnarray}
\label{D_split}
\mathcal{D} &=& \mathcal{D}_{\rm gh} + \mathcal{D}_V\;,\;\;\;
{\rm where}\;\;\;\mathcal{D}_{V} = \sum_{\alpha > 0} c^\alpha\,\mathcal{L}_{-\alpha}\;,\cr
\overline{\mathcal{D}} &=& \overline{\mathcal{D}}_{\rm gh} + 
\overline{\mathcal{D}}_V\;,\;\;\;
{\rm where}\;\;\;
\overline{\mathcal{D}}_{V} = \sum_{\alpha > 0} c^{-\alpha}\,\mathcal{L}_{\alpha}\;.
\end{eqnarray}
Using the $^\dagger$-conjugate of the identities (\ref{j_plus_de}): $\mathcal{D}^\dagger 
= - [\,J^-\,,\,\overline{\mathcal{D}}\,]$ and $\overline{\mathcal{D}}^\dagger 
=  [\,J^-\,,\,\mathcal{D}\,]$  one may calculate directly:
	\begin{eqnarray}
	\label{D_split_dagger}
	\mathcal{D}_V^\dagger &=& - \sum_{\alpha>0}\frac{1}{r_\alpha} 
	b_{-\alpha}\,\mathcal{L}_{\alpha}\;\;\;
	{\rm and}\;\;\;\mathcal{D}_{\rm gh}^\dagger = - \sum_{\alpha>0}\frac{1}{r_\alpha} 
	b_{-\alpha}\,\mathcal{D}_{\alpha}\;,\\
	\label{D_bar_split_dagger}
	\overline{\mathcal{D}}_V^\dagger &=& - \sum_{\alpha>0}\frac{1}{r_\alpha} 
	b_{\alpha}\,\mathcal{L}_{-\alpha}\;\;\;
	{\rm and}\;\;\;\overline{\mathcal{D}}_{\rm gh}^\dagger = - \sum_{\alpha>0}\frac{1}{r_\alpha} 
	b_{\alpha}\,\mathcal{D}_{-\alpha}\;,
	\end{eqnarray}
where $\{\mathcal{D}_{\pm\alpha}\}_{\alpha>0}$ are the operators defined as follows:
	\begin{eqnarray}
	\label{D_alpha}
	\mathcal{D}_{\pm\alpha} &=& \tilde{L}_{\pm\alpha} + \Gamma_{\pm\alpha}\;\;\;
	{\rm with}\;\;\;\Gamma_{\pm\alpha} = \sum_{\beta >0}\frac{r_\alpha}{r_{\alpha+\beta}}\,
	N_{\beta\;-(\alpha+\beta)}\,c^{\mp\beta} b_{\pm(\alpha + \beta)} \;\;\;{\rm and}\cr\cr
	\tilde{L}_{+\alpha} &=& \{\,b_{\alpha} ,\overline{\mathcal{D}}_{\rm gh}\,\}\;,\;\;
	\tilde{L}_{-\alpha} = \{\,b_{-\alpha},\mathcal{D}_{\rm gh}\,\}\;,
	\end{eqnarray}
where, as usually, $\{\,\cdot,\cdot\,\}$ denotes the symmetric bracket. 
Taking into account the explicit formulae
\begin{equation}
\label{L_exp}
\tilde{L}_{\pm \alpha} = 
- \,\sum_{\beta >0}\, N_{\alpha\;\beta}\,c^{\mp \beta}\,b_{\pm(\alpha + \beta)} \,-\, 
\sum_{\beta > \alpha}\, N_{\alpha\;-\beta}\,c^{\pm \beta}\,b_{\pm(\alpha - \beta)}\,\;,
\end{equation}
  and the cocycle property of $r_{(\cdot)}$ one may find  the equivalent 
expresssion
\footnote{In fact the structure constants present in (\ref{D_alpha}) do satisfy
$ N_{\beta\;-(\alpha+\beta)} = - N_{\alpha\;\beta}$. From the point of view of the subsequent calculations it is however more convenient to use the expressions as they are displayed.}
 for the above elements, namely:
	\begin{equation}
	\label{alternative_exp}
	\mathcal{D}_{\pm\alpha} = t_{\pm\alpha} + \Theta_{\pm\alpha}\;\;\;
	{\rm where}\;\;\;
	\Theta_{\pm\alpha} = \sum_{\beta >0}\frac{r_\beta}{r_{\alpha+\beta}}\,
	N_{\beta\;-(\alpha+\beta)}\,c^{\mp\beta} b_{\pm(\alpha + \beta)}\;,
	\end{equation}
and where $\{t_{\pm\alpha}\}_{\alpha>0}$, already present on the right hand side of (\ref{L_exp}),  are defined in (\ref{d_bar}) and (\ref{d_nie_bar}). 
From (\ref{D_alpha}) or (\ref{alternative_exp}) it immediately follows that $\mathcal{D}_{-\alpha} = - \mathcal{D}_{\alpha}^\ast \,$.\\ 
It is worth to mention some additional remarkable and essential properties of the operators above. Using the last expression one may prove the simple statement on their conjugation rules with respect to $^\dagger$. They appear to be crucial for the properties of Laplace operators. In addition it is possible to say something on their commutation relations. \\
One has the following
 	\begin{lemma}\
\begin{enumerate}
\item
The operators  $\mathcal{D}_{\pm \alpha}$ do satisfy the following conjugation rules:
 \begin{equation}
	\label{D_conj}
{\mathcal{D}_{\pm \alpha}}^\dagger\, = 
	- \,\mathcal{D}_{\mp \alpha}\;.
	\end{equation}
\item
The operators $\mathcal{D}_{\pm \alpha}$ do satisfy the structural relations of $\,\mathfrak{n}_{\pm}$ nilpotent sublagebras:
	\begin{equation}
	\label{D_borel}
[\,\mathcal{D}_{\pm \alpha}\,,\,\mathcal{D}_{\pm \beta}\,] = 
	N_{\alpha\, \beta}\,\mathcal{D}_{\pm (\alpha+\beta)}\;.
	\end{equation}
\end{enumerate}
\end{lemma}
{\it Proof}:\\
{\it 1}.  The proof of the conjugation properties is straightforward if the expressions (\ref{alternative_exp}), the definitions (\ref{d_bar}), (\ref{d_nie_bar}) and (\ref{kahler_conjugation}) are taken into account. By direct computation with the help of cocycle identity (\ref{cocycle}) one shows that ${t_{\alpha}}^\dagger = - \Theta_{-\alpha}$ and ${\Theta_{\alpha}}^\dagger = - t_{-\alpha}\,$.\\
{\it 2}. The structural relations follow from the following observations. First of all all the operators kill the ghost vacuum (\ref{vacuum}): $\mathcal{D}_{\pm \alpha}\,\omega = 0\;,\;\alpha>0$. Hence in order to recover their commutation relations it is enough to identify their bracket action on the generators $b_{-\alpha}$ and $c^{-\alpha}$ of ghost vectors. By simple calculation one obtains: 
	$[\,\mathcal{D}_{- \alpha}\,,\,c^{-\beta}\,] = 
- N_{\alpha\;-(\alpha+\beta)}\, c^{-(\alpha+\beta)}$ and 
$[\,\mathcal{D}_{- \alpha}\,,\,b_{-\beta}\,] = 
- \frac{r_{\beta}}{r_{\alpha+\beta}}\,N_{\alpha\;-(\alpha+\beta)}\, b_{-(\alpha+\beta)}\,$. The first bracket can be easily recognized as (co)adjoint (remark of the footnote) action of $\mathfrak{b}_-$. The second one is the same after obvious change of basic generators: $b_{-\alpha} \rightarrow b_{-\alpha}/r_{\alpha}$. Hence the operators must satisfy the structural relations of $\mathfrak{b}_-$. The relations for $\mathcal{D}_{ \alpha}$ are obtained by $^\ast$ or $^\dagger$ conjugation. $\;\;\;\;\blacksquare$
\\
\\
The cross bracket relations of $\{\mathcal{D}_\alpha\}_{\alpha>0}$ with $\{\mathcal{D}_{-\beta}\}_{\beta>0}$ are quite complicated and far from being transparent. For this reason they are not displayed here.
\\
\\
Introducing the family of operators: 
\begin{equation}
\label{D_alpha_tot}
D_{\alpha} = \mathcal{D}_{\alpha} + \mathcal{L}_{\alpha}\;,\;\;\;\alpha \in R\,,
\end{equation} 
and using the formulae (\ref{D_split_dagger}) and (\ref{D_bar_split_dagger})  for $^\dagger$-conjugated differentials together with the property (\ref{D_conj}) one  obtains  the following expressions for differentials $\mathcal{D}$ and 
$\overline{\mathcal{D}}$:
\begin{equation}
\label{useful_exp}
\mathcal{D} = \sum_{\alpha>0}\, c^\alpha\,{D}_{-\alpha}\;,\;\;\;\;
\overline{\mathcal{D}} = \sum_{\alpha>0}\, c^{-\alpha}\,{D}_{\alpha}\;.
\end{equation}
\\
In order to prove the formulae (\ref{lap_str}) for Laplace operator $\Box$ one proceeds as follows.
The calculations are performed  in two almost separate moves. From (\ref{D_split}) and (\ref{D_split_dagger}) one obtains:
\begin{eqnarray}
\label{2_moves}
\Box &=& \Box_{\rm gh} + \Box_V\;,\;\;\;
\Box_{\rm gh} = \{\,\mathcal{D}_{\rm gh},{\mathcal{D}_{\rm gh}}^\dagger\,\} \cr\cr
\Box_V &=&  \{\,\mathcal{D}_{\rm gh}, {\mathcal{D}_{V}}^\dagger \,\}+ 
\{\,{\mathcal{D}_{\rm gh}}^\dagger, {\mathcal{D}_{V}}\,\} + 
\{\,{\mathcal{D}_{V}},{\mathcal{D}_{V}}^\dagger\,\}\;.
\end{eqnarray}
In the first step the appropriate expression for $\Box_V$ will be obtained. From the definition (\ref{2_moves}) one immediately obtains:
\begin{eqnarray}
\label{lap_V}
 \Box_V = &-& \sum_{\alpha\; \beta > 0} \frac{1}{r_{\beta}}
\left( c^\alpha b_{-\beta} \mathcal{L}_{-\alpha}\mathcal{L}_{\beta} +
b_{-\beta}c^\alpha  \mathcal{L}_{\beta}\mathcal{L}_{-\alpha}\right) \,-\,\\
\label{lap_V_1}
&-&\,\sum_{\beta >0}\,\frac{1}{r_{\beta}} \{\,\mathcal{D}_{\rm gh},b_{-\beta}\,\} 
\,\mathcal{L}_{\beta}\,+\,
\sum_{\alpha >0}\,\{\,c^\alpha ,{\mathcal{D}_{\rm gh}}^\dagger\,\}\,
\mathcal{L}_{-\alpha}\;.
\end{eqnarray}
With the help of ghost structural relations (\ref{clifford_r}) and Lie algebra structural relations of (\ref{action_com}) one may transform the operator of (\ref{lap_V}) to the following form:
\begin{eqnarray}
\label{lap_V_2}
( {\rm \ref{lap_V}}) = &-& \sum_{\alpha >0}\,\frac{1}{r_{\alpha}}\,
\mathcal{L}_{-\alpha}\,\mathcal{L}_{\alpha}\,+\,
\sum_{\alpha >0} \,\frac{1}{r_{\alpha}}\left( <\chi,H_{\alpha}> + \mathcal{L}_{H_{\alpha}}\right)\,b_{-\alpha}c^{\alpha}\\
\label{lap_V_3} 
&+& \sum_{\alpha\;\beta >0}\,\frac{1}{r_{\beta}} b_{-\beta}c^{\alpha} 
N_{-\alpha\;\beta}\,\mathcal{L}_{\beta - \alpha}\;.
\end{eqnarray}
The terms in (\ref{lap_V_2}) are already in convenient form. The remaining one  (\ref{lap_V_3}) will be combined with those of (\ref{lap_V_1}).\\
The first one denoted by (\ref{lap_V_1}.1) according to the definitions of (\ref{D_alpha}) gives:  
\begin{equation}
\label{lap_V_4}
{\rm (\ref{lap_V_1}.1)} = - \,\sum_{\alpha >0}\,\frac{1}{r_{\alpha}} \, \tilde{L}_{-\alpha}\,\mathcal{L}_{\alpha}\;.
\end{equation}
According to (\ref{D_split_dagger}) and again (\ref{D_alpha}) for the second one denoted by (\ref{lap_V_1}.2) one obtains:
\begin{equation}
\label{lap_V_5}
{\rm (\ref{lap_V_1}.2)} = - \,\sum_{\alpha >0}\,\frac{1}{r_{\alpha}} \, \mathcal{D}_{\alpha}\,\mathcal{L}_{- \alpha}\, + \,
\sum_{\alpha\;\beta >0}\,\frac{1}{r_{\alpha}} N_{\alpha\;-(\alpha+\beta)}\,
b_{-\alpha}\,c^{\alpha+\beta}\,\mathcal{L}_{-\alpha}                                                      \;.
\end{equation}
Now it will be shown  that the second sum above cancels the part of (\ref{lap_V_3}) containing the operators $\mathcal{L}_{\alpha}$ corresponding to positive roots. After straightforward change of the summation variables the sum (\ref{lap_V_3}) splits  
as follows:
\begin{equation}
\label{lap_V_6}
{\rm (\ref{lap_V_3})} = 
 - \sum_{\alpha\;\beta >0}\,\frac{1}{r_{\alpha}} N_{\alpha\;-(\alpha+\beta)}\,
b_{-\alpha}\,c^{\alpha+\beta}\,\mathcal{L}_{-\alpha} \;+\;
 \sum_{\alpha\;\beta >0}\,\frac{1}{r_{\alpha +\beta}} N_{\beta\;-(\alpha+\beta)}\,
b_{-(\alpha+\beta)}\,c^{\beta}\,\mathcal{L}_{\alpha}                                                   \;.
\end{equation}
The first term above cancels the last sum of (\ref{lap_V_5}). According to definition (\ref{D_alpha}) the last term above equals to:
\begin{equation}
\label{lap_V_7} 
 - \sum_{\alpha >0}\,\frac{1}{r_{\alpha}} 
\Gamma_{-\alpha}\,\mathcal{L}_{\alpha} \;,
\end{equation} 
and supplements (\ref{lap_V_4}) to give 
$\;- \sum_{\alpha >0} \frac{1}{r_{\alpha}} \mathcal{D}_{-\alpha}\,\mathcal{L}_{\alpha}\,$.\\
Hence finally one obtains
\begin{equation}
\label{lap_V_8} 
\Box_V = - \sum_{\alpha >0}\,\frac{1}{r_{\alpha}}\,\left(
\mathcal{L}_{-\alpha}\,\mathcal{L}_{\alpha}\,+
\mathcal{D}_{-\alpha}\,\mathcal{L}_{\alpha}\, +
\mathcal{D}_{\alpha}\,\mathcal{L}_{-\alpha}\right) \,+ \,
\sum_{\alpha >0} \,\frac{1}{r_{\alpha}}\left( <\chi,H_{\alpha}> + \mathcal{L}_{H_{\alpha}}\right)\,b_{-\alpha}c^{\alpha}\;.
\end{equation} 
\\
The calculation of the purely ghost part of the Laplace operator present in (\ref{2_moves})
is a bit more complicated technically. It is  in particular very important to extract from $\Box_{\rm gh}$ the terms which correspond to those present at the very end of the right hand side of (\ref{lap_V_8}). Together they will compose the appropriate ghost number operator $\rm{gh}$. On the other hand they will supplement the Cartan subalgebra elaments of (\ref{lap_V_8}) such that they become identically zero on relative cochains.\\
Using the formulae (\ref{D_split_dagger}) for the differential ${D_{\rm gh}}^\dagger$ expressed in terms of $\mathcal{D}_{\alpha}$ one 
gets\footnote{It is important to notice that 
$b_{\alpha}\mathcal{D}_{\alpha} = \mathcal{D}_{\alpha} b_{\alpha} \,$.}
\begin{equation}
\label{lap_gh_0}
- \Box_{\rm gh} = - \{\,\mathcal{D}_{\rm gh},{\mathcal{D}_{\rm gh}}^\dagger\,\} = 
\sum_{\alpha > 0}\,\frac{1}{r_{\alpha}}\,\tilde{L}_{-\alpha}\,\mathcal{D}_{\alpha} \,+\,
\sum_{\alpha > 0}\,\frac{1}{r_{\alpha}}\,b_{-\alpha}\,
[\,\mathcal{D}_{\alpha}\,,\,\mathcal{D}_{\rm gh}\,]\;,
\end{equation}
where $\{\tilde{L}_{-\alpha}\}_{\alpha >0}$ are defined by (\ref{D_alpha}) and are explicitely given in (\ref{L_exp}). Note that in order to obtain the desired result the summands under the first sum above must be additively supplemented by the terms of the form 
$\Gamma_{-\alpha} \,\mathcal{D}_{\alpha}\,$ analogous to those of (\ref{lap_V_7}). For this reason it is very important to extract them from the second sum on the right hand side of (\ref{lap_gh_0}) containing  the commutators. \\
In order to find  the missing ingredients of the Laplace operator  one uses the definition
(\ref{D_alpha}):
\begin{equation}
\label{lap_gh_1}
 [\,\mathcal{D}_{\alpha}\,,\mathcal{D}_{\rm gh}\,] = [\,\tilde{L}_{\alpha}\,,\mathcal{D}_{\rm gh}\,] \,+\,
[\,\Gamma_{\alpha}\,,\mathcal{D}_{\rm gh}\,]\;.
\end{equation} 
Taking into account that according to (\ref{D_alpha}) 
$\tilde{L}_{\alpha} = \{\,b_{\alpha}\,,\,\overline{\mathcal{D}}_{\rm gh}\,\}$ it is possible to exploit the graded Jacobi identities to obtain:
\begin{equation}
\label{lap_gh_2}
 [\,\tilde{L}_{\alpha}\,,\mathcal{D}_{\rm gh}\,] \, = \,
- \,[\,\{\,b_{\alpha},\mathcal{D}_{\rm gh}\,\}\,,\overline{\mathcal{D}}_{\rm gh}\,] \,+\,
[\,b_{\alpha}\,,\,\{\,\mathcal{D}_{\rm gh}\,,\,\overline{\mathcal{D}}_{\rm gh}\,\}]
\;.
\end{equation} 
It is now important to find the expression for the bracket $\{\,\mathcal{D}_{\rm gh}\,,\,\overline{\mathcal{D}}_{\rm gh}\,\}$ present in the formulae above. From (\ref{curvature}), the bigraded split formulae (\ref{d_bar}), (\ref{d_nie_bar}) and (\ref{dede_bar}) restricted to pure ghost sector together with the respective formulae for the ghost contribution to the square of relative differential one may deduce that:
\begin{equation}
\label{lap_gh_3}
 \{\,\mathcal{D}_{\rm gh}\,,\,\overline{\mathcal{D}}_{\rm gh}\,\} \,= \,
- \sum_{\alpha > 0}\,c^{-\alpha}\,c^{\alpha} \left(\,<2\varrho,H_{\alpha}> \,+\, L_{H_{\alpha}}\,\right)\;.
\end{equation} 
Hence the last term of (\ref{lap_gh_2}) inserted to the appropriate sum of (\ref{lap_gh_1}) gives:
\begin{equation}
\label{lap_gh_4}
- \sum_{\alpha > 0}\,\frac{1}{r_{\alpha}}\, \left(\,<2\varrho,H_{\alpha}> \,+\, L_{H_{\alpha}}\,\right)\,b_{-\alpha}\,c^{\alpha}\;.
\end{equation} 
It is now evident that the above term added to the last sum of (\ref{lap_V_8})  combines to the ghost number operator ${\rm gh} = \sum_{\alpha>0} \,b_{-\alpha}\,c^{\alpha}$. The terms containing the Cartan subalgebra elements $L_{H_{\alpha}} + \mathcal{L}_{H_{\alpha}}$ do vanish on the relative cochains. \\
The bracket $\{\,b_{\alpha} ,\mathcal{D}_{\rm gh}\,\}$ present in (\ref{lap_gh_2}) will be  denoted by $\Upsilon_{\alpha}$. It can be easily found by direct computation:
\begin{equation}
\label{lap_gh_5}
\Upsilon_{\alpha} = \sum_{0<\beta <\alpha }\,N_{\beta\;-\alpha}\,c^{\beta}\,b_{\alpha - \beta} \;.
\end{equation} 
In order to calculate its commutator with $\overline{\mathcal{D}}_{\rm gh}$ it is convenient to use the expression (\ref{useful_exp}). One obtains:
\begin{equation}
\label{lap_gh_6}
[\,\Upsilon_{\alpha}, \overline{\mathcal{D}}_{\rm gh} \,]= \sum_{\beta > 0}\,[\,\Upsilon_{\alpha},c^{-\beta}]\,\mathcal{D}_{\beta} + 
\sum_{\beta >0 }\,c^{-\beta}\,[\,\Upsilon_{\alpha}, \mathcal{D}_{\beta}\,] \;.
\end{equation} 
The first term above inserted into the second sum of (\ref{lap_gh_0}) gives the missing summand:
$$
\sum_{\alpha>0}\,\frac{1}{r_{\alpha}}\,\Gamma_{-\alpha}\,\mathcal{D}_{\alpha}\;,
$$ 
which supplements the  first term of (\ref{lap_gh_0})  and gives the desired  sum 
$\sum_{\alpha>0}\frac{1}{r_{\alpha}}\mathcal{D}_{-\alpha}\mathcal{D}_{\alpha}\,$.
\\
\\
The considerations of this {\it Appendix} can be summarized as follows. It was demonstrated that adding up the operator $\Box_V$ of (\ref{lap_V_8}) and $\Box_{\rm gh}$ of (\ref{lap_gh_0}) one obtains:
\begin{equation}
\label{lap_gh_7}
\Box = -\sum_{\alpha > 0}\,\frac{1}{r_{\alpha}}\,D_{-\alpha}\,D_{\alpha} + {\rm gh}\;-\; 
\sum_{\alpha>0}\frac{1}{r_{\alpha}}\,b_{-\alpha}\,
(\,\sum_{\beta>0} c^{-\beta} [\,\Upsilon_{\alpha},\mathcal{D}_{\beta}\,] + 
[\,\Gamma_{\alpha}, \mathcal{D}\,]\, ) \;.
\end{equation} 
The authors strongly suspect that the last and ugly term in (\ref{lap_gh_7}) does vanish. The authors gave up however after unsuccesful attempts of calculations involving Jacobi identities expressed in terms of $N_{\alpha \beta}$ structure constants and cocycle property of $r_{\alpha}$ coefficients. For this reason the Proposition (\ref{lap_str}) remains with the status of conjecture.

\section*{Ackowledgements}
The authors would like thank dr Marcin Daszkiewicz for his inerest for the subject.  One of the authors (Z.H.) would like to thank his wife Meg for her understanding and patience.

\end{document}